\documentclass[12pt]{article}
\usepackage{amsmath,epsf,amssymb,latexsym,cite}
\usepackage[ps,dvips,matrix,arrow,frame,import,curve,color]{xy}

\newtheorem{theorem}{Theorem}

\newtheorem{conjecture}{Conjecture}

\newif\iffigs\figstrue

\DeclareFontFamily{U}{rsf}{}
\DeclareFontShape{U}{rsf}{m}{n}{
  <5> <6> rsfs5 <7> <8> <9> rsfs7 <10-> rsfs10}{}
\DeclareMathAlphabet\Scr{U}{rsf}{m}{n}

\def\pplogo{\vbox{\kern-\headheight\kern -29pt
\halign{##&##\hfil\cr&{
\ppnumber}\cr\rule{0pt}{2.5ex}&\ppdate\cr}
}}
\makeatletter
\def\ps@firstpage{\ps@empty \def\@oddhead{\hss\pplogo}%
  \let\@evenhead\@oddhead 
}
\def\maketitle{\par
 \begingroup
 \def\thefootnote{\fnsymbol{footnote}}
 \def\@makefnmark{\hbox{$^{\@thefnmark}$\hss}}
 \if@twocolumn
 \twocolumn[\@maketitle]
 \else \newpage
 \global\@topnum\z@ \@maketitle \fi\thispagestyle{firstpage}\@thanks
 \endgroup
 \setcounter{footnote}{0}
 \let\maketitle\relax
 \let\@maketitle\relax
 \gdef\@thanks{}\gdef\@author{}\gdef\@title{}\let\thanks\relax}
\makeatother

\def\O{\Scr{O}}
\def\C{{\mathbb C}}
\def\P{{\mathbb P}}

\def\R{{\mathbb R}}
\def\Z{{\mathbb Z}}

\def\Hom{\operatorname{Hom}}
\def\hom{\operatorname{hom}}

\def\Ext{\operatorname{Ext}}
\def\End{\operatorname{End}}

\def\Img{\operatorname{Im}}

\def\GU{\operatorname{U{}}}

\def\rank{\operatorname{rank}}

\def\Cone{\operatorname{Cone}}

\def\ch{\operatorname{\mathit{ch}}}
\def\td{\operatorname{\mathit{td}}}
\def\dP#1{\mathrm{dP}_{#1}}
\def\cmod#1{\hbox{$#1$--\bf mod}}

\def\p{\partial}

\def\sm{$\sigma$-model}
\def\nlsm{non-linear \sm}

\def\CY{Calabi--Yau}

\def\cE{{\Scr E}}
\def\cF{{\Scr F}}

\def\DC{\mathbf{D}}

\def\ff#1#2{{\textstyle\frac{#1}{#2}}}

\def\RHom{\mathbf{R}\Hom}
\def\poso#1{#1\save="x"!LD+<0pt,-0.5mm>;
  "x"!RD+<0pt,-0.5mm>**\dir{.}\restore}

\def\wPhi{\widetilde{\Phi}}
\def\wchi{\widetilde{\chi}}

\begin{document}
\setcounter{page}0
\def\ppnumber{\vbox{\baselineskip14pt
\hbox{DUKE-CGTP-04-05}
\hbox{hep-th/0405134}}}
\def\ppdate{May 2004} \date{}

\title{\LARGE D-Branes on Vanishing del Pezzo Surfaces\\[10mm]}
\author{
Paul S.~Aspinwall and Ilarion V.~Melnikov\\[3mm]
\normalsize Center for Geometry and Theoretical Physics \\
\normalsize Box 90318 \\
\normalsize Duke University \\
\normalsize Durham, NC 27708-0318
}

{\hfuzz=10cm\maketitle}

\def\Large{\large}
\def\LARGE{\large\bf}

\vskip 1cm

\begin{abstract}
We analyze in detail the case of a marginally stable D-Brane on a
collapsed del Pezzo surface in a \CY\ threefold using the derived
category of quiver representations and the idea of aligned
gradings. We show how the derived category approach to D-branes
provides a straight-forward and rigorous construction of quiver gauge
theories associated to such singularities. Our method shows that a
procedure involving exceptional collections used elsewhere in the
literature is only valid if some tachyon-inducing $\Ext^3$ groups are
zero. We then analyze in generality a large class of Seiberg dualities
which arise from tilting equivalences. It follows that some (but not
all) mutations of exceptional collections induce Seiberg duality in
this context. The same tilting equivalence can also be used to remove
unwanted $\Ext^3$ groups and convert an unphysical quiver into a
physical one.
\end{abstract}

\vfil\break


\section{Introduction}    \label{s:intro}

Suppose a D-brane is marginally stable against decay into a collection
of stable ``constituent'' or ``fractional'' D-branes. Each constituent
D-brane may appear with multiplicity $N_i$ and so is associated to a
factor of $\GU(N_i)$ in the world-volume gauge theory. The fact that
these D-branes are marginally bound implies that there are massless open
strings between them which correspond to chiral fields in
$(\overline{\mathbf{N}}_i,\mathbf{N}_j)$ representations in the above
gauge theory. In this way we associate a ``quiver'' gauge theory to
any D-brane decay.

For D-branes on a \CY\ threefold, we expect an enormous number
(probably dense) of walls of marginal stability in the moduli space
and so we should have a correspondingly huge number of possibilities for
quiver gauge theories.

The case most frequently studied concerns a BPS B-type D-brane
corresponding to a point on a \CY\ space $X$. If we place such a
D-brane at a ``singular'' point we expect to possibly find a marginal
decay. The best-understood case probably concerns orbifold
singularities locally of the form $\C^3/G$
\cite{DM:qiv,DDG:wrap,DFR:orbifold,Douglas:Dlect,me:TASI-D}. The
D-brane decays into a set of ``fractional branes'' associated to
irreducible representations of $G$ and the associated quiver is given
by the McKay quiver.

We would like to consider the case of a complex surface $S$ shrinking
down to a point inside $X$ to produce a singularity. Again, we would
expect the D-brane associated to this singular point to decay into
fractional branes. If $S$ is smooth and irreducible it must be a del
Pezzo surface. These cases overlap with the orbifolds only in the
single case that $S=\P^2$ corresponding to $\C^3/\Z_3$. Other del
Pezzo's do not produce orbifold singularities, and other orbifolds have
exceptional divisors with more than one component.

This case has been discussed many times in the literature
\cite{MP:AdS,BGLP:dPo,BP:toric,FHHI:quiv,HI:quiv,CFIKV:,
Wijn:dP,HW:dib,Herz:exc}. The essentially new thing we are going to do
in this paper is to bring the full weight of the machinery of the
derived category of coherent sheaves \cite{Doug:DC,AL:DC,me:TASI-D} to
bear on the problem so that we can make a clear statement (see theorem
\ref{th:main}). We believe this clarifies many aspects of this
subject.

An important ingredient in the derived category approach is an integer
grading. Given a D-brane $A$ one may produce another D-brane $A[n]$ by
``shifting $A$ $n$-places to the left.'' If $n$ is odd and $A$ is the
only D-brane under consideration then $A[n]$ is an ``anti''
D-brane. In other words we only care about $n$ mod 2. When we have
more than one D-brane the relative integer grading between branes
becomes important and one cannot simply reduce mod 2. That is, it is
too simplistic to talk in terms of branes and anti-branes. In this
paper we will see exactly how the latter picture can go wrong.

One can attack the problem we are interested in by using the A-model
description of the mirror. This may be done following the ideas of
\cite{HIV:D-mir,Sei:vanish,CFIKV:}. The mirror to the grading in the
derived category concerns the degree of Floer cohomology as given by
Maslov indices. Thus, reducing the grading mod 2 amounts to merely
counting points of intersection between 3-cycles, rather than the more
intricate procedure of computing the Floer cohomology groups. The
derived category picture of B-branes is generally much easier to
handle than Floer cohomology and the Fukaya category \cite{Fuk:cat} of
A-branes. It follows that mirror symmetry is not really a useful tool
given the degree of precision we desire this paper.

One of the most interesting aspects of quiver gauge theories concerns
``Seiberg dualities'' \cite{Sei:N1dual}. Although originally
considered as a thoroughly quantum effect in $N=1$ field theories, it
appears that these dualities correspond to equivalences between
D-branes even in the limit of zero string coupling. The derived
category picture of this story corresponds to ``tilting equivalences,''
as proposed by Berenstein and Douglas \cite{BD:tilt}. The main
obstacle to applying the full power of the mathematics of quiver
representations to Seiberg duality is the occurrence of oriented loops
in the quiver. If a quiver has an oriented loop, then the quiver
representations become infinite-dimensional, and it becomes much
harder to make specific statements (although an example was studied in
\cite{Brn:tilt}). Sadly, any physical quiver for the problems we are
analyzing in this paper has these unwanted oriented loops. 

In the case of del Pezzo surfaces, we may associate some of the
arrows in the gauge quiver to the intrinsic properties of the surface
itself, while the remaining arrows are associated to the embedding of
this surface into the \CY\ manifold. By deleting these latter arrows
we remove all oriented loops. Thus we are able to analyze the problem
using finite-dimensional representation theory. This is how we proceed
in this paper. Fortunately, it is a simple matter to add in the arrows
associated to the embedding after most of the analysis has been done.

This means we are able to provide a fairly general analysis of Seiberg
dualities for del Pezzo surfaces. An interesting fact that we will
observe is that there are two different tilting equivalences, each
the inverse of the other, associated to Seiberg duality. 
These tilts differ by whether the brane to anti-brane transformation is
given by a shift $[1]$ or a shift $[-1]$.  For many nodes in the
quiver only one these two tilts produces a valid duality.

In section \ref{s:marg} we review the general picture of how a
marginal decay yields a quiver gauge theory. Of particular interest is
the way in which this occurs because of an alignment in the gradings
of a large class of D-branes in the problem.

In section \ref{s:cat} we will review the mathematics of the derived
category of coherent sheaves on a del Pezzo surface and its relation
to quiver categories and exceptional collections of sheaves. This
involves the relationship between tilting complexes and projective
objects. We then show that the quiver associated to the derived
category of the del Pezzo surface is indeed the gauge quiver for an
object such as a marginally decaying 0-brane, so long as some $\Ext^3$
groups vanish.

In section \ref{s:tilt} we study how Seiberg dualities on the quiver
gauge theory are given by tilting equivalences. We then show how the
tilting picture can be tied in with the mutation picture for Seiberg
duality which is used elsewhere in the literature. We also show how
tilts may also remove the troublesome $\Ext^3$'s induced by some
exceptional collections.

When this paper was completed, a paper appeared with a large overlap 
with the work presented here \cite{Herz:ouch}.


\section{Marginal Decays and Quivers} \label{s:marg}

Suppose we have a B-type BPS D-brane, $A$, on a \CY\ manifold $X$. Let
us assume it fills the spatial directions of uncompactified
spacetime. Our convention will be to use the notation ``$n$-brane''
where $n$ refers to the dimensionality of the brane {\em within the
\CY\/}. Thus a 0-brane, as we denote it, would correspond to a point
on $X$ and would be considered a 3-brane in the full ten-dimensional
spacetime.

It is well-established
\cite{Doug:DC,AL:DC,Douglas:Dlect,me:TASI-D,KS:Ext}, that the B-type
topological D-branes on $X$ are objects in $\DC(X)$, the derived
category of coherent sheaves on $X$. In order to correspond to a
physical D-brane, such an object must be $\Pi$-stable
\cite{DFR:stab,Doug:DC,AD:Dstab}. This $\Pi$-stability condition
depends both on the complexified K\"ahler form, $B+iJ$, of $X$ and
upon the position (i.e., moduli) of the D-brane in $X$.

Of particular interest will be the case where the D-brane is a 0-brane
corresponding to a point $p\in X$. Generally speaking one expects such
a D-brane to be stable if $p$ corresponds to a smooth point. If $p$ is
at a singularity one might expect the 0-brane to decay.

The general picture for $\Pi$-stability proceeds as follows. Each
stable B-brane has a ``grade'' $\xi\in\R$ which varies with $B+iJ$. 
The grading is defined mod 2 by the argument of the central charge:
\begin{equation}
  \xi(A) = \frac1\pi\arg Z(A) \pmod2.
\end{equation}
The mod 2 ambiguity can be fixed from the large radius limit
\cite{me:TASI-D}. $\xi$ is not a single-valued function of the K\"ahler
moduli but should be thought of as a function on the Teichm\"uller
space of $B+iJ$. The Hilbert space of scalar objects in
the D-brane world-volume corresponding to open strings from a B-brane $B$
to another B-Brane $A$ in the sector seen by the topological field theory is
given by $\oplus_p\Ext^p(A,B)$. The mass of such an open string is
given by \cite{Doug:DC}
\begin{equation}
  m^2 = \ff12(\xi(A)-\xi(B)+p-1).   \label{eq:mass}
\end{equation}
If this string is tachyonic, then it will bind $A$ to $B$. If it is
massive, then, in the absence of any other bindings, $A$ and $B$ will
not be bound. Thus if we move in the moduli space by varying $B+iJ$, the
gradings $\xi$ of each B-brane will vary and the spectrum of stable
B-branes will jump as the masses of the above open strings pass
through zero.

Suppose we are in a situation where the grades, $\xi$, of a set of
B-branes $L_0,L_1,\ldots,L_{n-1}$ coincide. We will discuss examples
of this shortly. Let us also assume the following relation is
satisfied:
\begin{equation}
  \Hom(L_i,L_j) = \C\delta_{ij}. \label{eq:noHom}
\end{equation}
Since $\Hom=\Ext^0$, according to (\ref{eq:mass}), there are no
tachyons betweens these D-branes.\footnote{The tachyon from $L_i$ to
  itself is removed by the GSO projection.} The map from $L_i$ to itself gives
rise to a vector particle and thus (classically) a $\GU(1)$ gauge theory in the
D-brane world-volume. As usual, if we have $N_i$ B-branes of the type
$L_i$, we would obtain a $\GU(N_i)$ gauge theory.

According to (\ref{eq:mass}) the massless scalars between $L_i$ and
$L_j$ would be counted by $\Ext^1(L_i,L_j)$ and $\Ext^1(L_j,L_i)$. If
we have $N_i$ copies of each $L_i$, we would have $\Ext^1(L_i,L_j)$
scalar fields transforming in a $(\overline{\mathbf{N}}_i,\mathbf{N}_j)$
representation of the $\GU(N_i)\times\GU(N_j)$ part of the gauge
group.

Therefore, when the grades of a set of B-branes coincide and satisfy
(\ref{eq:noHom}), we automatically have a ``quiver gauge
theory.'' Each node in the quiver is labeled by $i$ and corresponds to
a $\GU(N_i)$ factor of the gauge group. Each arrow in the diagram
corresponds to an $\Ext^1$ group and is interpreted as a bifundamental
chiral field in the four-dimensional D-brane world volume theory.

This quiver gauge theory can be thought of as describing a marginal
binding of the associated D-branes. The bifundamental chiral fields
are exactly massless. A perturbation of $B+iJ$ may make these open
strings tachyonic or massive, making the bound state stable or unstable,
respectively.

The open strings $\Ext^p(L_i,L_j)$ for $p>1$ correspond to very
massive strings and will be ignored in the quiver gauge theory.

We would like to study marginally-stable B-branes corresponding to
isolated points in a \CY\ threefold. Presumably such an instability
requires the point be a singularity. Since the stability is governed
by $B+iJ$, the singularity must be obtained by a deformation of
$B+iJ$, i.e., a blow-down of something inside the \CY\ threefold. If
the subspace that is blown-down to a point is a smooth irreducible
surface $S$, then this surface must be a del Pezzo surface
\cite{Reid:can3}. We will restrict attention to this case in this
paper.

The case of $S=\P^2$ collapsing to a point corresponds to the
$\C^3/\Z_3$ orbifold. The relevant analysis for this problem has been
studied in
\cite{AGM:sd,DFR:orbifold,DDG:wrap,DG:fracM,me:TASI-D}. There it is
established that the three fractional branes into which the 0-brane
decays indeed have exactly the same value for their grading when
precisely at the orbifold point in the moduli space. Thus our picture
applies. Any perturbation away from this point in the moduli space
would destroy the marginal decay of the 0-brane into the 3 fractional
branes. At least one bound state of two of the three fractional branes
would be definitely stable or unstable.

The del Pezzo surfaces are $\P^1 \times \P^1$ and $\P^2$ blown-up at $m$
points, which we denote by $\dP m$. There are thus 
$\dim H^{\textrm{even}}=m+3$ periods characterizing the central charges
for branes on $\dP m$. There is an irrelevant overall
scaling of the periods, so we have $m+2$ independent periods. The
moduli space of $B+iJ$ has complex dimension $\dim H^2=m+1$. Thus, in order
to align the gradings of all the B-branes, i.e., align all the
$\arg$'s of the periods, we would need to impose $m+2$ real
constraints on $2m+2$ real moduli. Thus we expect a subspace of
the moduli space of real dimension $m$ for which the gradings are suitably
aligned. This ties in with the case $m=0$ (the orbifold above), where
there was only one point in the moduli space.

Our expectation is therefore that the alignment occurs somewhere in
moduli space where the del Pezzo surface has shrunk to zero size and
there may be remaining degrees of freedom in the moduli (given by
$B$-fields) which do not effect this alignment. We show this
explicitly for $\P^1\times\P^1$ in the next section.

In a way, this alignment has already been demonstrated for any
singularity that can be obtained by the partial resolution of an
orbifold singularity. The McKay correspondence proves that gradings
align at the orbifold point (theorem 4 of \cite{me:TASI-D}). A partial
resolution of this orbifold should preserve the alignment of the fractional
branes associated with the remaining singularity. This covers a wide
class of del Pezzo surfaces. It would be nice to make this argument
rigorous.

\subsection{Gradings and Periods for $S = \P^1 \times \P^1$}

In this section we find the subspace in the moduli space where the
gradings are aligned for the example of $S = \P^1 \times \P^1.$ As
discussed in \cite{me:TASI-D}, the central charges of the B-branes on
$X$ are given by a period computation on the mirror of $X$.  Which
periods appear is determined by matching the form of the D-brane
charge in the large radius limit:
\begin{equation}
\label{centralchargeX}
 Z(\cE) = \int_X e^{-(B+iJ)} \ch(\cE) \sqrt{\td(X)}.
\end{equation}
We will be concerned with branes on $S \subset X$.  Let
$\eta_1,\eta_2$ be elements in $H^2(S,\Z)$ dual to the hyperplane
classes of the $\P^1$'s.  Expressing $B+iJ$ in this dual basis
as $B+iJ= t_1 \eta_1 + t_2 \eta_2,$ $Z(\cE)$ reduces to
\begin{equation}
\label{centralchargeS}
 Z(\cE) = \int_S e^{-(t_1 \eta_1 + t_2 \eta_2)} 
\ch(\cE) \sqrt{\frac{\td(S)}{\td(N_S)}},
\end{equation}
where $N_S$ denotes the normal bundle to $S$.  In the large radius
limit ($\Img(t_1), \Img(t_2) \rightarrow \infty$) where the above
expression is valid, we find that $Z(\cE)$ has terms that scale as
$1$, $t_1$, $t_2$, and $t_1 t_2$.  The relevant subspace of the moduli
space of the \nlsm ~on $X$ is most conveniently parametrized by the
B-model coordinates $z_1, z_2,$ which are related to $t_1,t_2$ by the
mirror map.  Asymptotically, this is given by
\begin{eqnarray}
t_1 \sim \frac{1}{2\pi i} \log(z_1), \nonumber\\
t_2 \sim \frac{1}{2\pi i} \log(z_2). 
\end{eqnarray}
More generally, $t_1,t_2$ are given by ratios of periods on the
mirror.  It is clear that $Z(\cE)$ will be expressible as a linear
combination of four periods that in the large radius limit ($\left| z_1
\right|, \left| z_2 \right| \to 0$) scale as
\begin{eqnarray}
\Phi_0 & \sim & 1, \nonumber\\
\Phi_1 & \sim &  \frac{1}{2 \pi i} \log(z_1), \nonumber\\
\Phi_2 & \sim &  \frac{1}{2 \pi i} \log(z_2), \nonumber \\
\Phi_3 & \sim &  -\frac{1}{4 \pi^2} \log(z_1)\log(z_2).
\end{eqnarray}

The gradings will align for those $(z_1,z_2)$ where these periods are
simultaneously real.  To find these points in the moduli space, we
must first calculate these periods.  We do this by finding solutions
to the Picard-Fuchs equations with appropriate asymptotics.  In the
case of toric $S$, it is known that the Picard-Fuchs system is a
special case of the GKZ system, and for $S = \P^1 \times \P^1,$ the
Picard-Fuchs system is given by
\begin{eqnarray}
 \left( \theta_1^2 - 4 z_1 
\left(\theta_1 + \theta_2 + 1\right) 
\left(\theta_1+\theta_2\right) \right) \Phi(z_1,z_2) = 0, \nonumber \\
\left( \theta_2^2 - 4 z_2 \left(\theta_1 + \theta_2 + 1\right) 
 \left(\theta_1+\theta_2\right) \right) \Phi(z_1,z_2) = 0, 
\end{eqnarray}
where $\theta_a = z_a \frac{\p}{\p z_a}.$ 

By using the techniques of \cite{AGM:sd,me:min-d} we find the
following solutions:
\begin{eqnarray}
\Phi_0(z_1,z_2) & = & 1, \nonumber\\
\Phi_1(z_1,z_2) & = & \frac{1}{2\pi i} \log(e^{i \pi} z_1) + 
\frac{1}{i \pi} \sum_{(m,n)\neq(0,0)} A_{mn} z_1^m z_2^n, \nonumber \\
\Phi_2(z_1,z_2) & = & \frac{1}{2\pi i} \log(e^{i \pi} z_2) + 
\frac{1}{i \pi} \sum_{(m,n)\neq(0,0)} A_{mn} z_1^m z_2^n, \nonumber \\
\Phi_3(z_1,z_2) & = & -\frac{1}{4\pi^2} \Biggl(\log(e^{i\pi} z_1) 
\log(e^{i\pi} z_2)\\
 &&\qquad+ \sum_{(m,n)\neq(0,0)} A_{mn} z_1^m z_2^n 
                        \left( \log(e^{i\pi} z_1) + 
\log(e^{i\pi} z_2) + \chi_{mn} \right)\Biggr),\nonumber
\end{eqnarray}
where
\begin{eqnarray}
A_{mn}    & = & \frac{\Gamma(2m + 2n)}{\Gamma(m+1)^2 
\Gamma(n+1)^2}, \nonumber \\
\chi_{mn} & = & 4 \Psi(2m+2n) - 2 \Psi(m+1)  -2 \Psi(n+1),
\end{eqnarray}
$\Psi(z)$ is the usual di-Gamma function, and the sum
$\sum_{(m,n)\neq(0,0)}$ runs over all non-negative $m,n$ with the
exception of $m=n=0$. The power series converge for $
\left|z_1\right| < \ff14$ and $\left|z_2\right| < \ff14$. These radii of
convergence are determined by the distance of the large radius limit
point to the discriminant locus.

Having found the appropriate solutions in the large radius limit, we
can analytically continue these periods to a phase where $S$ shrinks
by using Mellin-Barnes representations of these solutions.  We
continue to a phase where $\left|z_2\right| \rightarrow \infty,$ and
the appropriate coordinates are given by $y_1 = z_1/z_2,$ and $y_2 =
1/z_2.$ Denoting by $\wPhi_i$ the analytic continuation of $\Phi_i,$
we find
\begin{eqnarray}
\wPhi_0(y_1,y_2)  & = & 1, \nonumber\\
\wPhi_1(y_1,y_2)  & = & \frac{1}{2\pi i} \log(y_1) + \wPhi_2(y_1,y_2), 
\nonumber\\
\wPhi_2(y_1,y_2)  & = &-\frac{1}{2\pi i} 
\left(e^{-i\pi}y_2\right)^{\frac{1}{2}}
                       \sum_{m,n\geq0} B_{mn} y_1^m y_2^n, \nonumber\\
\wPhi_3(y_1,y_2)  & = & \frac{1}{12} - \frac{1}{4\pi^2} 
\left(e^{-i \pi} y_2\right)^{\frac{1}{2}}
                       \sum_{m,n\geq0} B_{mn} y_1^m y_2^n 
\left(-\log(y_1) + \wchi_{mn} \right),
\end{eqnarray}
where
\begin{eqnarray}
B_{mn}     & = & \frac{\Gamma(m+n+\frac{1}{2})^2}{\pi 
\Gamma(m+1)^2 \Gamma(n+1)^2}, \nonumber \\
\wchi_{mn} & = & 2 \Psi(m+1) - 2 \Psi(m+n + \ff12).
\end{eqnarray}
It is clear that in order for $\wPhi_1$ and $\wPhi_2$ to be
simultaneously real, we must have $\left|y_1\right| = 1.$ Setting $y_2
= 0,$ we find that the periods are simultaneously real.  Thus, as
expected from the dimension counting given above, we have found a
one-dimensional real subspace where the gradings align.

The reader may be worried that $\left|y_1\right| = 1$ is right on the
radius of convergence of the power series in $\wPhi_i.$ One can show
that the series converge for $y_1 \neq 1.$ The divergence at $y_1 = 1$
is easy to understand: $y_1=1, y_2 = 0$ is on the discriminant locus
of this model, which is given by
\begin{equation}
P(y_1,y_2) = 16 y_1^2 - 8 y_1 y_2 + y_2^2 - 32 y_1 -8 y_2 + 16.
\end{equation}

Unlike the case of the $\P^2$, here the grading align near the
discriminant locus and not at the orbifold point.  This is an
important difference that is likely to persist for other del Pezzo
surfaces.  The example of $\P^1\times \P^1$ is particularly tractable
due to the symmetry between $z_1$ and $z_2.$ The other toric del Pezzo
surfaces can be treated in much the same fashion as above, but the
computations are quite a bit more involved.


\section{The Category of B-branes on del Pezzo Surfaces} \label{s:cat}

\subsection{Quivers and Algebras}  \label{ss:quiv}

Before we can attack the problem of del Pezzo surfaces, we require some
knowledge of the mathematics of quivers and tilting. See
\cite{Douglas:Dlect} and references therein for an account of the way
quivers first appeared in the context at hand. We refer to
\cite{KZ:tilt} for a complete description of tilting. See also
\cite{BD:tilt,Brn:tilt} for accounts in the physics literature.

The first ingredient we require is the concept of the path algebra of
a quiver. Let $Q$ be a quiver with nodes $v_i$ and arrows
$a_\alpha$. The path algebra $A$ of $Q$ is generated as follows. To each
node $v_i$ we associate an element $e_i$ considered to be a path of
length 0. The other generators consist of nonzero-length paths in the
quiver. Clearly, each arrow $a_\alpha$ may be associated to a path. If
the head of $a_\alpha$ is the same node as the tail of $a_\beta$, then
we may produce a path $a_\beta a_\alpha$ consisting of $a_\alpha$
followed by $a_\beta$. Note that we compose paths right-to-left in our
notation. This order is very important.

Multiplication in $A$ is then defined as composition of paths in
the obvious way. If the end of a path $\gamma_1$ is not the same node
as the start of a path $\gamma_2$ then we define
$\gamma_2\gamma_1=0$. Note that the zero-length paths $e_i$ are
idempotent: $e_i^2=e_i$.

We can also impose {\em relations\/} on the quiver by asserting some
relations the paths must obey. This amounts to setting $A$ equal to
some algebra generated by the paths divided by an ideal generated by
the relations. For example, consider the following quiver
\begin{equation}
\begin{xy} <1.0mm,0mm>:
  (0,0)*{\circ}="a",(20,0)*{\circ}="b",(40,0)*{\circ}="c",
  (0,-4)*{v_0},(20,-4)*{v_1},(40,-4)*{v_2}
  \ar@{<-}@/^3mm/|{a_0} "a";"b"
  \ar@{<-}|{a_1} "a";"b"
  \ar@{<-}@/_3mm/|{a_2} "a";"b"
  \ar@{<-}@/^3mm/|{b_0} "b";"c"
  \ar@{<-}|{b_1} "b";"c"
  \ar@{<-}@/_3mm/|{b_2} "b";"c"
\end{xy}  \label{eq:eg1}
\end{equation}
We could choose to (and will) impose the 3 relations 
$a_\alpha b_\beta=a_\beta b_\alpha$.

Let $V$ be a given representation of the algebra $A$ or, equivalently,
a left $A$-module. Using the idempotent elements $e_i$, we form vector
spaces $V_i=e_iV$. Let $N_i=\dim(V_i)$.  The elements of $A$
corresponding to arrows in $Q$ then correspond to linear maps between
the $V_i$'s. That is, $V$ can be associated to a set of integers
$n_i$, one for each node, and a set of matrices, one for each
arrow. Clearly these matrices will have to satisfy any relations that
have been imposed on the quiver.  This latter data is a ``quiver
representation.'' The inverse of this procedure is easily constructed
(see \cite{Ben:quiv}, for example), showing that representations of $A$
are equivalent to quiver representations of $Q$.

We may define a {\em morphism\/} between two representations $W$ and
$V$ of $A$ as a linear map $\phi:W\to V$ such that $a\phi(w)=\phi
a(W)$ for any $w\in W$ and $a\in A$. Translating this into the
language of quiver representations, this amounts to a set of linear
maps $\phi_i:W_i\to V_i$ such that the $\phi_i's$ commute with the
maps within each quiver in the obvious way. Using the quiver
representations as objects and the above morphisms, one defines the
category of representations of $Q$ (or equivalently left
$A$-modules). 

As a simple and useful example of a short exact sequence
of quiver representations consider \def\biS#1#2{
\begin{xy} <0.8mm,0mm>:
  (0,0)*{\circ};(0,10)*{\circ}**\dir{-}
    ?(0.58)*\dir{>}, (0,-3)*{\scriptstyle#1}, (0,13)*{\scriptstyle#2},
  (-5,0)*i{x},(5,0)*i{x}
\end{xy}}
\begin{equation}
\xymatrix@1{
0\ar[r]&\biS01\ar[r]&\biS11\ar[r]&\biS10\ar[r]&0.\POS(36.5,0)*{\scriptstyle
  f}} \label{eq:extQ}
\end{equation}
The numbers in this diagram represent the dimensions $n_i$ and $f$ is
multiplication by any complex number. The horizontal maps in this
sequence between nontrivial vector spaces need not be zero. However,
if $f$ is nonzero, there are no nonzero morphisms going in the reverse
directions to the ones shown.

There are two distinguished sets of useful quiver representation
associated to a given quiver $Q$. In each case they are labeled by
the nodes $i$. The first obvious set, $L_i$, corresponds to the
one-dimensional representations given by $n_j=\delta_{ij}$. The second
set, $P_i$, is defined by $P_i=A e_i$. That is $P_i$ is the subspace of
$A$ generated by all paths {\em starting\/} at node $i$. Multiplying
on the left by elements of $A$ makes $P_i$ a left $A$-module and thus
a representation.

Using $(N_0,N_1,\ldots)$ to denote the dimensions of representations,
where $N_i=\dim(V_i)$ as above, it is easy to see in the example
(\ref{eq:eg1}) that the dimensions are as follows
\begin{equation}
\begin{split}
  \dim L_0 &= (1,0,0),\\
  \dim L_1 &= (0,1,0),\\
  \dim L_2 &= (0,0,1),\\
  \dim P_0 &= (1,0,0),\\
  \dim P_1 &= (3,1,0),\\
  \dim P_2 &= (6,3,1).
\end{split}   \label{eq:lp2}
\end{equation}
One can show that the $P_i$ are {\em projective\/} objects in the
category of representations.

Note that if the quiver has any directed loops then some of the
$P_i$'s will be infinite-dimensional. This makes the analysis of such
quivers considerably more difficult and most of the methods used in
this paper will be useless. Luckily, by restricting our
attention to del Pezzo surfaces we will effectively evade this
case. From now on we will assume that there are {\em no\/} directed loops 
in the quiver $Q$.

Given that there are no directed loops in the quiver, we may assign an
order to the labels of the nodes. We will assert that there is no path
from node $i$ to node $j$ if $i<j$. This is consistent with our
example (\ref{eq:eg1}).

It is not hard to show that the $P_i$'s form a complete set of
projective objects in the sense that any quiver representation has a
{\em projective resolution\/} in terms of sums of $P_i$'s. By using
(\ref{eq:lp2}), one can see in our example that the following exact
sequences form projective resolutions of the $L_i$'s:
\begin{equation}
\xymatrix@R=3mm{  
&&0\ar[r]&P_0\ar[r]&L_0\ar[r]&0\\
&0\ar[r]&P_0^{\oplus3}\ar[r]&P_1\ar[r]&L_1\ar[r]&0\\
0\ar[r]&P_0^{\oplus3}\ar[r]&P_1^{\oplus3}\ar[r]&P_2\ar[r]&L_2\ar[r]&0\\
}
\end{equation}
In general we will write the projective resolutions of the $L_i$
representations as
\begin{equation}
\xymatrix@1{
\ldots\ar[r]&\bigoplus_{k}P_k^{\oplus r_{ik}}
  \ar[r]&\bigoplus_{k}P_k^{\oplus n_{ik}}\ar[r]&P_i\ar[r]&L_i\ar[r]&0.
}  \label{eq:Lres}
\end{equation}
One can show that $n_{ij}$ is equal to the number of arrows in the
quiver from node $i$ to node $j$ and that $r_{ij}$ represents the
number of independent relations imposed on paths from $i$ to
$j$. Another basic fact about these representations which is easily
proven is that
\begin{equation}
  \Hom(P_i,L_j) = \delta_{ij}\C.   \label{eq:homPL}
\end{equation}

We would now like to compute some $\Ext$ groups which are central to
our analysis. This is very easy in the current context. If an object $A$
has a projective resolution
\begin{equation}
\xymatrix@1{
\ldots\ar[r]&\Pi_2\ar[r]&\Pi_1\ar[r]&\Pi_0\ar[r]&A\ar[r]&0,
}  \label{eq:Pres}
\end{equation}
where the $\Pi_i$ are projective objects (and thus direct sums of
$P_i$'s), then $\Ext^p(A,B)$ is given by the cohomology of the complex
\begin{equation}
\xymatrix@1{
0\ar[r]&\Hom(\Pi_0,B)\ar[r]&\Hom(\Pi_1,B)\ar[r]&\Hom(\Pi_2,B)
\ar[r]&\ldots
}
\end{equation}
in the $p$th position.

Thus we may use the resolutions
(\ref{eq:Lres}) together with (\ref{eq:homPL}) to compute 
\begin{equation}
\begin{split}
  \dim\Ext^1(L_i,L_j) &= n_{ij}\\
  \dim\Ext^2(L_i,L_j) &= r_{ij}.
\end{split}
\end{equation}
One should therefore think of the arrows in a quiver as representing
$\Ext^1$'s between the basic $L_i$ representations and $\Ext^2$'s as
arising because of relations in the quiver. Note that viewing the 
short exact sequence (\ref{eq:extQ}) provides another way of seeing
that the arrows in a quiver correspond to $\Ext^1$'s.

The ordering we have chosen on the nodes implies that there can never
be a non-zero map in the resolution $P_i\to P_j$ if $i>j$. This
implies that $\Ext^p(L_i,L_j)=0$ for any $p$ if $i<j$.

We may now make contact with the quiver gauge theories of section
\ref{s:marg}. The arrows in the quiver gauge theory correspond to
bifundamental chiral fields and are counted by $\Ext^1$'s between the
D-branes in the derived category. {\em Therefore we associate the
fractional D-branes at the nodes of quiver with the basic
representations $L_i$.} The precise description for how composite
D-branes decay into these fractional branes will be given in section
\ref{ss:decay}.

\subsection{Tilting}  \label{ss:tilt}

Given a quiver $Q$ with path algebra $A$, let us denote the category
of quiver representations (or left $A$-modules) by $\cmod A$. We may
now define the derived category\footnote{All derived categories in
this paper are bounded.}  $\DC(\cmod A)$ to be the derived category
obtained by passing to complexes of quiver representations. We refer
the reader to \cite{me:TASI-D} for what is intended to be a relatively
gentle introduction to derived categories. Given another quiver with a
path algebra $B$ we would like to know when $\DC(\cmod A)$ is
equivalent to $\DC(\cmod B)$.

If $Z$ is a quiver representation, i.e., an object in $\cmod A$, we
will also use $Z$ to denote an object in $\DC(\cmod A)$ consisting of
a complex whose only nonzero entry is $Z$ at position zero.  With this
notation in mind, consider a quiver $Q$ with $n$ nodes and define the
object
\begin{equation}
  T = P_0\oplus P_1\oplus\ldots\oplus P_{n-1},  \label{eq:tilt0}
\end{equation}
where the $P_i$'s are the projective objects from section
\ref{ss:quiv}, and let
\begin{equation}
  C = \End(T),
\end{equation}
i.e., $C$ is the algebra of morphisms of $T$ back to
itself. Multiplication in this algebra is simply composition of
morphisms. To each $P_i$ we can clearly associate an idempotent
element $e_i$ in $C$ corresponding to the projection of $T$ onto
$P_i$. Following carefully through the definition of morphisms above
one can see that $\Hom(P_i,P_j)$ is given by the vector space of paths
from $j$ to $i$. This means that $C$ is exactly the algebra one would
associate to the quiver $Q$ {\em if all the arrows were reversed}.

We may define the algebra $\End(T)^{\textrm{op}}$ to be that given by
$\End(T)$ except that the order of composition is reversed. This has
the effect of reversing the direction of morphisms and thus we regain
the original quiver. That is,
\begin{equation}
  A \cong \End(T)^{\textrm{op}}.  \label{eq:Atilt0}
\end{equation}

The idea of ``tilting'' is to replace $T$ in (\ref{eq:tilt0}) by a
more general direct sum of objects satisfying particular conditions in
order to get a new path algebra, distinct from $A$, but which is
identical as far as derived categories are concerned. 

To be precise, define a {\em tilting complex\/} $T$ to be an object in
$\DC(\cmod A)$ such that
\begin{enumerate}
\item $\Hom(T,T[i])=0$ for $i\neq0$.
\item The direct summands of $T$ can be used to generate the whole of 
$\DC(\cmod A)$ by translations and mapping cones.
\end{enumerate}
Here, as usual, we use the notation $T[i]$ to mean a left-shift of $T$
by $i$ places.\footnote{We will also, as usual, define $\Ext^p(A,B)$
to be $\Hom(A,B[p])$.} Mapping cones are the
natural way of combining objects via a morphism in the derived
category and correspond to combining D-branes via tachyon
condensation. Again we refer to \cite{me:TASI-D} for a review of this.

We then have following theorem due to Rickard \cite{Rick:der}
(following work by Happel \cite{Happ:der1})
\begin{theorem}
The derived categories $\DC(\cmod A)$ and $\DC(\cmod B)$ are
equivalent if and only if there exists a tilting complex
$T$ such that $B=\End(T)^{\textrm{op}}$.
\end{theorem}

The tilting complex given by (\ref{eq:tilt0}) clearly gives the
equivalence of $\DC(\cmod A)$ to itself. It satisfies condition ``1.''\ as a
tilting complex since $\Ext^p(P_i,P_j)=0$ for $p>0$ and condition ``2.''\
since all objects have a projective resolution.

\subsection{Del Pezzo Surfaces}   \label{ss:delP}

Let $S$ be a del Pezzo surface, i.e. a smooth surface whose
anticanonical class intersects every algebraic curve in $S$ a positive
number of times, and let $i:S\to X$ be the embedding of this surface
into a \CY\ threefold $X$.  Given an object in $\DC(S)$, we may use
the functor $i_*$ to map this object into a D-brane in
$\DC(X)$. Physically this is the obvious statement that a B-brane in $S$
may be viewed as a B-brane in $X$ if $S$ is embedded in $X$.

In order to describe B-branes wrapping on $S$, we need to describe
the derived category $\DC(S)$. Fortunately this is a well-known problem
in algebraic geometry, and there are some very powerful tools
established. In particular we may use the machinery of exceptional
collections of sheaves (see \cite{Rud:Ebook} are references therein)
based on Beilinson's \cite{Bei:res} construction for $\P^n$. 

Let $\{\cF_0,\ldots,\cF_{n-1}\}$ be an {\em exceptional
  collection\/} of sheaves on $S$. That is
\begin{equation}
\begin{split}
\Ext_S^p(\cF_i,\cF_i) &= \begin{cases}
  \C&\text{if $p=0$},\\
  0&\text{otherwise}, \end{cases}\\
\Ext_S^p(\cF_i,\cF_j) &= 0 \quad\text{for any $p$ and $i>j$}.
\end{split}
\end{equation}

One can then prove \cite{KO:excdP} that, if $i<j$, then
$\Ext^p(\cF_i,\cF_j)$ is nonzero for at most one value of $p$. A
collection of sheaves is said to be {\em strongly\/} exceptional if
$\Ext^p_S(\cF_i,\cF_j)=0$ for $p\neq0$. That is, only
$\Hom_S(\cF_i,\cF_j)$ can be nonzero. An exceptional collection is
said to be {\em complete\/} if it generates $\DC(S)$. This latter
condition is equivalent \cite{KO:excdP} to the number of elements in
the exceptional collection being equal to the Euler characteristic of
$S$.

If $S$ is a del Pezzo surface $\dP m$ with exceptional curves $C_1,
C_2, \ldots C_m$, then a strongly exceptional collection is given by
$\{\O,\O(C_1),\O(C_2),\ldots,\O(C_m),\O(H),\O(2H)\}$.  Here $H$ is a
hyperplane $\P^1$ not intersecting any of the $C_i$'s. Any exceptional
collection may be obtained from this one by a sequence of mutations
\cite{KO:excdP}. We will discuss mutations in section \ref{ss:mut}.

Assume $\{\cF_0,\ldots,\cF_{n-1}\}$ form a complete strongly exceptional
set of sheaves on $S$ and define
\begin{equation}
  A = \End(\cF_0\oplus \cF_1\oplus\ldots\oplus
  \cF_{n-1})^{\textrm{op}}.  \label{eq:Aex}
\end{equation}
Bondal \cite{Bon:dPq} then proved\footnote{Bondal refers to {\em
    right\/} modules but these are turned into left modules by the
    ``op'' in (\ref{eq:Aex}).}
\begin{theorem}
  The derived category of coherent sheaves $\DC(S)$ on $S$ is
  equivalent to the derived category $\DC(\cmod A)$.
\end{theorem}

Comparing this to (\ref{eq:Atilt0}) shows that the $\cF_i$'s are
playing the same r\^ole as the $P_i$'s of section \ref{ss:tilt}. In
other words, the derived category of a del Pezzo surface $S$ is equivalent
to the derived category of representations of a quiver $Q$ where the
projective representations $P_i$ correspond to a strongly exceptional
set of sheaves.

This means that, given a strongly exceptional set of sheaves on $S$,
we may construct the quiver immediately (and the set of relations)
since we know that $\Hom_S(\cF_i,\cF_j)$ is precisely the space of
paths from node $j$ to node $i$. For example, consider the strongly
exceptional collection $\{\O,\O(1),\O(2)\}$ on $S=\P^2$. Both
$\Hom(\O,\O(1))$ and $\Hom(\O(1),\O(2))$ are given by $\C^3$ and
correspond to multiplication by the homogeneous coordinates on
$\P^2$. $\Hom(\O,\O(2))\cong\C^6$ and is given by homogeneous
quadratic function of the homogeneous coordinates. Any element of
$\Hom(\O,\O(2))$ is given by an element of $\Hom(\O,\O(1))$ composed
with an element of $\Hom(\O(1),\O(2))$, and thus  no extra
arrows are needed between node 0 and node 2. In addition, we have an obvious
relation $x_iy_j=x_jy_i$ for the composition of such maps. Thus, after
reversing the arrows in accord with the above description, we see that
the quiver corresponding to $\P^2$ is given by our earlier example
(\ref{eq:eg1}).

As another example, consider $S=\dP1$ given by $\P^2$ with the single
point $[z_0,z_1,z_2]=[0,0,1]$ blown up. Using the strongly exceptional
collection $\{\O,\O(C_1),\O(H),\O(2H)\}$, the corresponding quiver is
given by
\begin{equation}
\begin{xy} <1.0mm,0mm>:
  (0,0)*{\circ}="a",(20,0)*{\circ}="b",(40,0)*{\circ}="c",(60,0)*{\circ}="d",
  (0,-4)*{v_0},(20,-4)*{v_1},(40,-4)*{v_2},(60,-4)*{v_3}
  \ar@{<-}|{a} "a";"b"
  \ar@{<-}@/^3mm/|{b_0} "b";"c"
  \ar@{<-}|{b_1} "b";"c"
  \ar@{<-}@/_8mm/|{c} "a";"c"
  \ar@{<-}@/^3mm/|{d_0} "c";"d"
  \ar@{<-}|{d_1} "c";"d"
  \ar@{<-}@/_3mm/|{d_2} "c";"d"
\end{xy}  \label{eq:dP1}
\end{equation}
subject to the relations $b_0d_1-b_1d_0=0$, $ab_0d_2-cd_0=0$, and
$ab_1d_2-cd_1=0$.

We emphasize that the sheaves in the exceptional collection are the
projective objects $P_i$ and not the fractional branes $L_i$. The
relationship between these two sets of D-branes was given by
projective resolutions (\ref{eq:Lres}). In references such as
\cite{CFIKV:,Wijn:dP} the exceptional sheaves themselves were taken to be the
fractional branes. In \cite{Herz:exc} the sheaves corresponding to
$L_i$ were called a ``dual collection'' to the given collection $P_i$.

The physical problem we wish to analyze concerns D-branes on $S$
embedded in $X$. Thus we need to apply the $i_*$ map discussed at the
start of this section. Objects in $\DC(S)$ are mapped injectively to
objects in $\DC(X)$, however, there may be more
morphisms between two objects in $\DC(X)$ than there were in
$\DC(S)$. In other words, there are some open string states between
two D-branes on $S$ that live ``outside'' $S$ in the threefold $X$.

Given two objects $A$ and $B$ in $\DC(S)$, we may use a spectral
sequence to compute the full spectrum of open strings
$\Ext^m_X(i_*A,i_*B)$ as used in \cite{me:point,KS:Ext}, for
example. Let $N$ be the normal bundle to $S$ in $X$. Then we have
spectral sequence with
\begin{equation}
  E_2^{p,q} = \Ext^p_S(A,B\otimes \wedge^q N)
\end{equation}
converging to $\Ext^{p+q}_X(i_*A,i_*B)$. In our case, because $X$ is a
\CY\ manifold, $N$ is equal to $K_S$, the canonical line bundle of
$S$. Furthermore, Serre duality tells us that $\Ext^p_S(A,B\otimes K_S)=
\Ext^{2-p}_S(B,A)$. The $E_2$ stage of our spectral sequence therefore
looks like
\begin{equation}
\begin{xy}
\xymatrix@C=4mm@R=3mm{
  \ldots&0&0&0&0&\ldots\\
  \ldots&\Ext^3_S(B,A)&\Ext^2_S(B,A)&\Ext^1_S(B,A)&\Ext^0_S(B,A)&\ldots\\
  \ldots&\Ext^{-1}_S(A,B)&\Ext^0_S(A,B)&\Ext^1_S(A,B)&\Ext^2_S(A,B)&\ldots\\
} 
\save="x"!LD+<-3mm,0pt>;"x"!RD+<0pt,0pt>**\dir{-}?>*\dir{>}\restore
\save="x"!LD+<37mm,-3mm>;"x"!LU+<37mm,-2mm>**\dir{-}?>*\dir{>}\restore
\save!CD+<0mm,-4mm>*{p}\restore
\save!UL+<34mm,-5mm>*{q}\restore
\end{xy}  \label{eq:StoX}
\end{equation}
We have allowed for $\Ext^p$'s with $p<0$ since we are working in the
derived category.

Because we have no directed loops in the quiver associated to a del
Pezzo surface, we may order the nodes and thus, as observed in section
\ref{ss:delP}, either $\Ext^p(L_i,L_j)$ or $\Ext^p(L_j,L_i)$ must
vanish for all $p$ (except in the trivial case that $i=j$). Hence, only
one row in (\ref{eq:StoX}) can contain nonzero entries, which immediately
implies that there can be no $d_2$ or higher differentials in the
spectral sequence. That is, the spectral sequence degenerates
immediately to yield
\begin{equation}
  \Ext^p_X(i_*L_i,i_*L_j) = \Ext^p_S(L_i,L_j) \oplus
  \Ext^{3-p}_S(L_j,L_i).
     \label{eq:ExtX}
\end{equation}

Actually there is a even stronger vanishing statement for the $\Ext$'s
appearing in the spectral sequence. One can argue (see \cite{HW:dib}
for example) that the $L_i$'s may be obtained from the given
exceptional collection by a sequence of mutations. Corollary 2.11 of
\cite{KO:excdP} states that $\Ext^p(L_i,L_j)$ will then be
nonzero for at most one value of $p$. Thus, in the case that $A$ and
$B$ are distinct fractional branes, at most one term in the diagram
(\ref{eq:StoX}) is nonzero.

The gauge quiver for the del Pezzo has arrows corresponding to
$\Ext^1$'s. Therefore, according to (\ref{eq:ExtX}) we need to add
arrows to the quiver corresponding to $\Ext^2_S(L_j,L_i)$ from node
$i$ to node $j$ to account for the extra open strings induced by the
embedding of $S$ in $X$. As seen in section \ref{ss:quiv} these extra
arrows are counted by the number of relations. Given an initial
quiver $Q$, we will refer to the new quiver with the added arrows as
the ``completed quiver'' and denote it $\bar Q$. The completed quiver
is the gauge quiver.

The completed quiver for $\P^2$ therefore becomes
\begin{equation}
\begin{xy} <1.0mm,0mm>:
  (0,0)*{\circ}="a",(20,0)*{\circ}="b",(40,0)*{\circ}="c",
  (0,-4)*{v_0},(20,-4)*{v_1},(40,-4)*{v_2}
  \ar@{-}@/_1mm/|*\dir{<}"a";"b"
  \ar@{-}|*\dir{<}"a";"b"
  \ar@{-}@/^1mm/|*\dir{<}"a";"b"
  \ar@{-}@/_1mm/|*\dir{<}"b";"c"
  \ar@{-}|*\dir{<}"b";"c"
  \ar@{-}@/^1mm/|*\dir{<}"b";"c"
  \ar@{.}@/^4mm/|*\dir{>}"a";"c"
  \ar@{.}@/^5mm/|*\dir{>}"a";"c"
  \ar@{.}@/^6mm/|*\dir{>}"a";"c"
\end{xy}  \label{eq:eg1X}
\end{equation}
in agreement with the McKay quiver of $\C^3/\Z_3$, and the quiver for
a $\dP1$ becomes
\begin{equation}
\begin{xy} <1.0mm,0mm>:
  (0,0)*{\circ}="a",(20,0)*{\circ}="b",(40,0)*{\circ}="c",(60,0)*{\circ}="d",
  (0,-4)*{v_0},(20,-4)*{v_1},(40,-4)*{v_2},(60,-4)*{v_3}
  \ar@{-}|*\dir{<}"a";"b"
  \ar@{-}@/_1mm/|*\dir{<}"b";"c"
  \ar@{-}|*\dir{<}"b";"c"
  \ar@{-}@/_6.5mm/|(0.3)*\dir{<}"a";"c"
  \ar@{-}@/_1mm/|*\dir{<}"c";"d"
  \ar@{-}|*\dir{<}"c";"d"
  \ar@{-}@/^1mm/|*\dir{<}"c";"d"
  \ar@{.}@/^6mm/|*\dir{<}"a";"d"
  \ar@{.}@/^5mm/|*\dir{<}"a";"d"
  \ar@{.}@/^4mm/|(0.3)*\dir{<}"b";"d"
\end{xy}  \label{eq:dP1X}
\end{equation}
We will always use dotted arrows to represent the new arrows added in.

The completed quiver, once the extra arrows have been added in,
contains oriented loops. This is always the
case.\footnote{~(\ref{eq:ExtX}) implies that
$\Ext^0_X(i_*L_j,i_*L_j)=\Ext^3_X(i_*L_j,i_*L_j)=\C$ for all $j$. The
projective resolution of $i_*L_j$ must therefore contain projective
objects $P_j$ in both the zero position and the third position ruling
out any ordering of the nodes. Indeed, every node must be contained in
at least one oriented loop.}  Because of this it is much harder to
analyze the gauge quiver directly using techniques of quiver
representations. The great thing about del Pezzo surfaces is that they
allow us to consider a subcategory $\DC(S)$ for which there are no
oriented loops. Almost all of the time when we analyze quivers in this
paper we will be treating the simpler loop-free quiver associated to
$\DC(S)$.

Equation (\ref{eq:ExtX}) also shows us that $\Ext^3_S(B,A)$ will
contribute to $\Ext^0_X(i_*A,i_*B)$. This is undesirable since
$\Ext^0$'s correspond to tachyons by (\ref{eq:mass}). In other words,
if $\Ext^3(L_i,L_j)$ is nonzero for any pair of fractional branes then
$L_i$ and $L_j$ will form a strongly bound state and completely
rearrange the quiver in question. Thus, the prescription for
determining the quiver from an exceptional collection
breaks down.

There are cases where this problem occurs. For example, a strong
complete exceptional collection on a $\dP4$ is given by
$\{\O,\O(C_1),\O(C_2),\O(C_3),\O(C_4),\O(H),\O(2H)\}$. The quiver is
then
\begin{equation}
\begin{xy} <1.0mm,0mm>:
  (0,0)*{\circ}="a",
  (20,15)*{\circ}="b1",
  (20,5)*{\circ}="b2",
  (20,-5)*{\circ}="b3",
  (20,-15)*{\circ}="b4",
  (40,0)*{\circ}="c",(60,0)*{\circ}="d",
  (0,-4)*{v_0},(20,11)*{v_1},(20,1)*{v_2},(20,-9)*{v_3},(20,-19)*{v_4},
  (40,-4)*{v_5},(60,-4)*{v_6}
  \ar@{-}|*\dir{<}"a";"b1"
  \ar@{-}|*\dir{<}"a";"b2"
  \ar@{-}|*\dir{<}"a";"b3"
  \ar@{-}|*\dir{<}"a";"b4"
  \ar@{-}@/^1mm/|*\dir{<}"b1";"c"
  \ar@{-}@/_1mm/|*\dir{<}"b1";"c"
  \ar@{-}@/^1mm/|*\dir{<}"b2";"c"
  \ar@{-}@/_1mm/|*\dir{<}"b2";"c"
  \ar@{-}@/^1mm/|*\dir{<}"b3";"c"
  \ar@{-}@/_1mm/|*\dir{<}"b3";"c"
  \ar@{-}@/^1mm/|*\dir{<}"b4";"c"
  \ar@{-}@/_1mm/|*\dir{<}"b4";"c"
  \ar@{-}@/^1.5mm/|*\dir{<}"c";"d"
  \ar@{-}|*\dir{<}"c";"d"
  \ar@{-}@/_1.5mm/|*\dir{<}"c";"d"
\end{xy}  \label{eq:dP4}
\end{equation}
The projective resolution of $L_6$ is
\begin{equation}
\xymatrix@1{
  0\ar[r]&P_0\ar[r]&P_1\oplus P_2\oplus P_3\oplus P_4\ar[r]&
   P_5^{\oplus3}\ar[r]&P_6\ar[r]&L_6\ar[r]&0,
}
\end{equation}
and thus $\Ext^3(L_6,L_0)=\C$. The appearance of $\Ext^3$'s may also be
viewed as ``relations between the relations'' in the quiver.

One way of evading these dangerous $\Ext^3$'s is to restrict attention
to the ``three block exceptional collections'' of \cite{KNog:3b}, as
was done in \cite{Wijn:dP}. This guarantees that no projective
resolution of $L_i$ need be more than 3 terms long, and thus the
problematic $\Ext^3$'s are always zero. We should note, however, that
the motivation for using three block exceptional collections given in
\cite{Wijn:dP}, concerning canceling charges and anti-branes, does not
apply when the derived category description is properly taken into
account. There are also perfectly good non-three-block cases such as
the $\dP1$ example we analyzed above.

It is perhaps conceivable that $\Ext^p(L_i,L_j)$ could be nonzero for
$p\geq4$ inducing further tachyons. We have not looked for such
examples. In summary, given any strongly exceptional complete
collection of sheaves on a del Pezzo surface we may construct the
associated quiver gauge theory of fractional branes by the above
methods if and only if $\Ext^p(L_i,L_j)=0$ for all $p>2$ and all pairs
of fractional branes $L_i$ and $L_j$.\footnote{If, and only if, these
conditions apply, then it follows that the intersection matrix
inversion trick of dual exceptional collections as used in
\cite{HW:dib,Herz:exc} will also yield the gauge quiver.}

\subsection{Decay into Fractional Branes}  \label{ss:decay}

Suppose $E$ is a quiver representation of dimension
$(N_0,N_1,\ldots,N_{n-1})$. We may also view $E$ as an object in the
derived category $\DC(S)$ as explained above.

Recall that, since there are no directed loops
in the quiver associated to a del Pezzo surface, we may number the
nodes such that there is no path from node $i$ to node $j$ if $i<j$.
Once we number this way, we may construct the following set of
distinguished triangles in $\DC(S)$:
\begin{equation}
\xymatrix@!C=3.5mm{
**[l]L_{0}^{\oplus N_{0}}=E_{0}\ar[rr]&&E_{1}\ar[dl]
\ar[rr]&&\cdots\ar[dl]\ar[rr]&&E_{n-2}\ar[dl]
\ar[rr]&&**[r]E_{n-1}=E\ar[dl]\\
&L_{1}^{\oplus N_{1}}\ar[ul]|{[1]}&&L_{2}^{\oplus N_{2}}\ar[ul]|{[1]}
&\ldots&L_{n-2}^{\oplus N_{n-2}}\ar[ul]|{[1]}&&L_{n-1}^{\oplus N_{n-1}}
\ar[ul]|{[1]}
} \label{eq:bchain}
\end{equation}
This chain also appeared in the work of Bridgeland \cite{Brg:stab}
and was discussed in the context of D-brane decay in
\cite{me:TASI-D}. We now show that this chain of distinguished
triangles exhibits how $E$ decays into the collection of fractional
branes $L_0^{\oplus N_0}\oplus L_1^{\oplus N_1}\oplus \ldots \oplus
L_{n-1}^{\oplus N_{n-1}}$. The object $E_k$ is a quiver representation
with dimension $(N_0,N_1,\ldots,N_{k},0,0,\ldots)$.

If, as we assume, the gradings of all the $L_i$ are equal, then by the
rules of $\Pi$-stability \cite{AD:Dstab}, $L_{0}^{\oplus N_{0}}$
and $L_{1}^{\oplus N_1}$ are marginally bound in the left-most
triangle forming $E_{1}$ with the same grading $\xi$. This in turn
implies that $E_{1}$ and $L_{2}^{\oplus N_{2}}$ are marginally
bound forming $E_{2}$. We continue iteratively along the chain of
triangles and see that $E$ is a marginally bound state of the $L_i$'s
as desired.

This is the last ingredient we required to prove the following
theorem:
\begin{theorem}
Suppose we are in a location in the $B+iJ$ moduli space corresponding
to a collapsed del Pezzo surface $S$ and where the gradings of the
basis of fractional branes are aligned. Then the decay of composite
D-branes is associated to a quiver gauge theory, where the quiver is
given by the completion of a quiver associated to the path algebra
$\End(T)^{\textrm{op}}$ and $T$ is the sum of a strong complete
exceptional collection of sheaves on $S$.  The fractional branes are
associated to the one-dimensional quiver representations $L_i$, where
$i$ labels the nodes in the quiver. The quiver gauge theory is
tachyon-free if and only if $\Ext^p(L_i,L_j)=0$ for all $i$ and $j$
and $p\geq3$.
\label{th:main}
\end{theorem}

The above discussion of decay has taken place in the context of
$\DC(S)$. What happens when we embed $S$ into $X$? $E$ is now
associated to a quiver representation with extra arrows associated
to $\Ext^2$'s. If this quiver representation associates nonzero
matrices to any of these new arrows, then we will kill some of the
morphisms used to construct the decay chain (\ref{eq:bchain}). If, on
the other hand, we associate a zero matrix with each new arrow then
nothing changes in our analysis above.

In other words, any quiver representation which associates nonzero
matrices to the ``new'' arrows coming from $\Ext^2$'s, will not be
marginally stable against a decay into the fractional branes. The
interpretation is clear --- turning on the matrices associated to the
new arrows created by the embedding corresponds to moving the
D-brane $E$ away from $S$ inside $X$. This was already understood in
the case $S=\P^2$ \cite{DFR:orbifold}.

The decay chain (\ref{eq:bchain}) allows one to compute the dimensions
of a quiver representation of a 0-brane as follows. We may apply the
functor $\Hom(P_k,-)$ to each of the distinguished triangles in the
chain. From (\ref{eq:homPL}), starting from the left in the decay
chain, this yields $\Hom(P_k,E_0)=0=
\Hom(P_k,E_1)=\ldots=\Hom(P_k,E_{k-1})=0$. The next triangle gives
$\Hom(P_k,E_k)=\C^{N_k}$. Continuing to the right finally yields
\begin{equation}
  \Hom(P_k,E) = \C^{N_k}.
\end{equation}

In our case, the projective objects $P_k$ are given by exceptional
sheaves $\cF_k$. If $E$ is a 0-brane then $E=\O_p$, the sky-scraper
sheaf of a point. In this case $\Hom(P_k,E)=\Hom(\cF_k,\O_p)$, where
the latter is given by the rank of the sheaf $\cF_k$. Thus we obtain
\begin{equation}
  N_k = \rank(\cF_k).
  \label{eq:Nrank}
\end{equation}
In the examples above, the exceptional collections were comprised only
of line bundles. The dimension of the quiver representation
corresponding to a 0-brane is therefore $(1,1,\ldots,1)$ in each case.

Since $\O_p\otimes K_S=\O_p$, Serre duality tells us that
$\Ext^n(A,E)\cong\Ext^{2-n}(E,A)$ for any $A$ if $E$ is the
0-brane. Defining, as usual
\begin{equation}
  \chi_X(\cE,\cF) = \sum_i (-1)\dim\Ext^i_X(\cE,\cF),
\end{equation}
this implies, from (\ref{eq:StoX}), that $\chi_X(i_*A,i_*E)=0$ for any
object $A$ in $\DC(S)$. It follows \cite{CFIKV:,Wijn:dP} that
\begin{equation}
  \sum_i N_i(\bar n_{ik}-\bar n_{ki}) = 0,\quad\hbox{for all $k$},
      \label{eq:anom1}
\end{equation}
where $\bar n_{ik}$ is the number of arrows from node $i$ to node $k$ in
the {\em completed\/} gauge quiver (dotted arrows and all). This
guarantees certain anomaly cancellations for the world-volume theory of the
0-brane \cite{CFIKV:}.

In this paper we are principally concerned with marginal
stability. However, one might also wish to know about the stability of
$E$ if we move away from our point of marginally stability. In this
case, to leading oder, $\Pi$-stability becomes equivalent to
the $\theta$-stability of \cite{King:th} as argued in
\cite{DFR:stab}. Given the decay chain (\ref{eq:bchain}) one can argue
\cite{me:TASI-D} that the following may be deduced from the ordering of
the gradings:
\begin{itemize}
\item $\xi(L_1)>\xi(L_2)>\ldots>\xi(L_{n-1})$ is a necessary and
  sufficient condition that $E$ decays completely into $L_1\oplus
  L_2\oplus\ldots\oplus L_{n-1}$.
\item If $E$ is a single-term complex and the maps within the
  quiver are sufficiently generic (such as for $\O_x$),
  $\xi(L_1)<\xi(L_2)<\ldots<\xi(L_{n-1})$ is a sufficient (but not
  necessary) condition that $E$ is stable.
\end{itemize}


\section{Tilting Duality}   \label{s:tilt}

\subsection{A class of tilts}  \label{ss:refl}

In section \ref{ss:tilt} we considered the tilting procedure which
allows one to demonstrate the equivalence of the derived category for
different quivers. Actually we only considered the identity tilt of a
quiver back to itself. We would now like to consider nontrivial tilts
which, as observed in \cite{BD:tilt}, correspond to Seiberg duality
\cite{Sei:N1dual}. These tilts are similar to ones analyzed in 
\cite{BGP:coxeter}.\footnote{Except that \cite{BGP:coxeter} considers
only nodes where the arrows are all incoming or all outgoing.}

Define the following tilting complex, denoted $T_L$, for a general
quiver with no oriented cycles and $n$ nodes. We choose a particular
node $k$ such that at least one arrow has its tail on this node, i.e.,
$n_{kj}\neq 0$ for some $j$.
\begin{equation}
  T_L = \bigoplus_{j=0}^{n-1} P'_j,  \label{eq:TS1}
\end{equation}
where
\begin{equation}
\begin{split}
  P'_j &= P_j\quad\hbox{for $j\neq k$,}\\
  P'_k &= \bigl(\xymatrix@1{\poso{\bigoplus_j P_j^{\oplus n_{kj}}}
     \ar[r]&P_k}\bigr),
\end{split}
\end{equation}
where the dotted underline means position zero in the complex. The
morphism in the complex for $P'_k$ is given by the natural composition
of paths and, in particular, is nonzero.

$T_L$ satisfies condition ``2.'' for a tilting complex in section
\ref{ss:tilt} since $P_k$ can be obtained from $P_k'$ from cone
constructions to cancel out the added $P_j$'s. Condition ``1.'' is
harder to prove. Let us introduce a little helpful notation. Let $C$
be an object in the derived category given by a chain complex whose
entries are $C^i$. Given two objects $C$ and $D$ in the derived
category we may produce $\hom(C,D)$ which is a complex of vector
spaces with each entry given by
\begin{equation}
  \hom^i(C,D) = \oplus_j\Hom(C^j,D^{j+i}),
\end{equation}
and obvious differential maps in the complex. We refer to
\cite{ST:braid} for more details. The cohomology of this complex
in the $i$th position is then $\Ext^i(C,D)$.

First note that, for $i\neq k$,
\begin{equation}
\begin{split}
  \hom(P_i',P_k') &= \bigl(\xymatrix@1{\poso{\bigoplus_j\Hom(P_i,P_j)^{\oplus
  n_{kj}}}\ar[r]&\Hom(P_i,P_k)}\bigr)\\
  &=\bigl(\xymatrix@1{\poso{\bigoplus_j(\hbox{paths $j\to
  i$})^{\hbox{\#(arrows
  $k\to j$)}}}\ar[r]&(\hbox{paths $k\to i$})}\bigr),
\end{split}
\end{equation}
where ``(paths $j\to i$)'' means the vector space generated by such paths.
Since all paths from $k$ to $i$ must pass through a $j$ in the sum, it
is easy to see that the map in the latter complex is surjective. This
implies that $\Ext^p(P_i',P_k')=0$ for $p\neq0$.  Similarly
\begin{equation}
\begin{split}
  \hom(P_k',P_i') &= \bigl(\xymatrix@1{\poso{\bigoplus_j\Hom(P_j,P_i)^{\oplus
  n_{kj}}}&\ar[l]_-f\Hom(P_k,P_i)}\bigr)\\
  &=\bigl(\xymatrix@1{\poso{\bigoplus_j(\hbox{paths $i\to
  j$})^{\hbox{\#(arrows
  $k\to j$)}}}&\ar[l]_-f(\hbox{paths $i\to k$})}\bigr)
\end{split} \label{eq:hom2}
\end{equation}
This map $f$ (going from right to left) in the latter complex is given as
follows. Given a path from $i$ to $k$, compose it with each arrow from
$k$ to $j$ to obtain a path from $i$ to $j$. If this map were not
injective it would imply that there is a nonzero linear combination of
paths from $i$ to $k$ that, when composed with any nontrivial path
starting at $k$, gives zero. Note that since we imposed $n_{kj}\neq0$
for some $j$, there certainly are nontrivial paths starting at node
$k$. 

If the map $f$ in (\ref{eq:hom2}) is injective we will call the object
$T_L$ ``admissible'' (following the language of \cite{Bon:quiv}). In
this case $\Ext^p(P_k',P_i')=0$ for $p\neq0$. It is easy to see in
the case that all the objects are line bundles that $T_L$ must be
admissible. In the general higher rank case, this need no longer be
true. The theory of mutations in section \ref{ss:mut} can be used to
prove that $f$ must be either injective or surjective. These cases may
be distinguished simply by computing the dimensions of the vector
spaces involved. Thus, it is a simple matter to determine if a given
$T_L$ is admissible.

Finally, one can also show that $\hom(P_k',P_k')=\C$. If $T_L$ is
admissible, then combining the above results yields
$\Hom(T_L,T_L[i])=0$ if $i\neq0$, and thus $T_L$ is a tilting complex.

The tilting recipe then tells us that we can construct an algebra 
$B=\End(T_L)^{\textrm{op}}$. This is the path algebra of a new quiver
$Q'$ which has the same derived category as the original quiver. What
is the new quiver? We need to compute $\Ext^p(L_i',L_j')$ where the
$L_i$'s are the new fractional branes for the quiver $Q'$.

In order to identify the $L_i'$'s we require more details of how the
tilting process yields an equivalence of derived categories. This is
provided by a functor $\Psi_L:\DC(\cmod A)\to\DC(\cmod B)$. If $C$ is an
object in $\DC(\cmod A)$, it can be shown that \cite{KZ:tilt}
\begin{equation}
  \Psi_L(C) = \RHom(T_L,C).
\end{equation}
If $E$ is an $A$-module, we can see that $\Hom_A(T_L,E)$ is a $B$-module as
follows. Let $h$ be an endomorphism of $T_L$ and let $f$ be a map from
$T_L$ to $E$. Then $h(f)$ is simply $f\circ h$.

Given (\ref{eq:TS1}), the new idempotent maps $e'_i$ for the path
algebra $B$ of the quiver $Q'$ are clearly projections of $T_L$ to
$P'_i$. Thus, if the dimension of $\Psi_L(E)$ is $(N_0',N_1',\ldots,
N_{n-1}')$, then
\begin{equation}
\begin{split}
  N_i' &= \dim(e'_i\Psi_L(E))\\
       &= \dim\Hom(P_i',E).
\end{split} \label{eq:Ni}
\end{equation}
Here we have assumed that $\Psi_L(E)$ is a quiver representation rather
than a complex of such representations, but the generalization to the
derived category involves the usual manipulations.

Choose a node $j$ of $Q$ such that there are no arrows from $k$ to $j$, i.e.,
$n_{kj}=0$. From above it follows that $\Psi_L(L_j)$ has dimension given
by $N_i'=\delta_{ij}$. In other words $\Psi_L(L_j)=L_j'$.

Next consider node $k$. The only nonzero $\hom(P_i',L_k)$ is given by
\begin{equation}
  \hom(P_k',L_k) = \bigl(\xymatrix@1{\C\ar[r]&\poso{0}}\bigr).
\end{equation}
This implies that $\Psi_L(L_k)=L_k'[1]$.

\def\bind{\mathbin{\leftrightsquigarrow}} The remaining nodes in $Q$
are nodes $j$ such that $n_{kj}>0$. In this case it is easy to check
that the dimension of $\Psi_L(L_j)$ is the same as the dimension of
$L_j'\oplus L_k'^{\oplus n_{kj}}$. This does not mean that
$\Psi_L(L_j)$ really is isomorphic to $L_j'\oplus L_k'^{\oplus
n_{kj}}$. Some of the matrices associated with any arrows
between $j$ and $k$ in $Q'$ may be nonzero for $\Psi_L(L_j)$. We will
therefore use the more vague notation $L_j'\bind L_k'^{\oplus n_{kj}}$
for this object.  In summary
\begin{equation}
\Psi_L(L_j)=\begin{cases}
   L_k'[1]&j=k\\
   L_j'&j\neq k,\:n_{kj}=0\\
   L_j'\bind L_k'^{\oplus n_{kj}}&j\neq k\end{cases} \label{eq:LLp}
\end{equation}
Since the functor $\Psi_L$ is an equivalence of categories,
$\Ext^p(L_i,L_j)\cong\Ext^p(\Psi_L(L_i),\Psi_L(L_j))$. This, together with
(\ref{eq:LLp}), is sufficient for us to compute the $\Ext$ groups
between the $L_j'$s and thus the new quiver $Q'$. In particular, if
$n_{kj}=0$ and $j\neq k$ then
\begin{equation}
\begin{split}
  \Ext^p(L_j',L_k') &= \Ext^{p-1}(L_j,L_k)\\
  \Ext^p(L_k',L_j') &= \Ext^{p+1}(L_k,L_j)
\end{split}   \label{eq:tshf}
\end{equation}

Next assume $n_{kj}>0$ and hence $j<k$. Then $\Ext^p(L_j,L_k)=0$ for
any $p$ and so
$\Ext^{-1}(\Psi_L(L_j),\Psi_L(L_k))=\Hom(\Psi_L(L_j),L_k)=0$. Given that
$\Psi_L(L_j)$ is $L_j'\bind L_k'^{\oplus n_{kj}}$, the only way that
$\Hom(\Psi_L(L_j),L_k')$ can be zero is if there is at least one arrow
associated with a nonzero matrix from node $j$ to $k$ in $Q'$. There
are no directed loops in $Q'$ (since that would result in an
infinite-dimensional $B$) and so there are no arrows from $k$ to $j$
in $Q'$. It follows that there is a short exact sequence
\begin{equation}
\xymatrix@1{
0\ar[r]&L_k'^{\oplus n_{kj}}\ar[r]&\Psi_L(L_j)\ar[r]&L_j'\ar[r]&0.
}  \label{eq:Lbind}
\end{equation}
This gives a more precise description for $\Psi_L(L_j)$ than
$L_j'\bind L_k'^{\oplus n_{kj}}$.
Applying the functor $\hom(-,\Psi_L(L_k))$ to this sequence yields
\begin{equation}
\begin{split}
  \dim\Ext^1(L_j',L_k') &= n_{kj} \\
   &= \dim\Ext^1(L_k,L_j),
\end{split}
  \label{eq:flip}
\end{equation}
with all other $\Ext$ groups between $L_j'$ and $L_k'$ vanishing. This
sequence also implies that the short exact sequence (\ref{eq:Lbind})
is not split. This implies that the open string modes from $L_j'$ to
$L_k'^{\oplus n_{kj}}$ acquire nonzero values to form a ``bound
state'' $\Psi_L(L_k)$.

We would now like to compare this tilting equivalence with Seiberg
duality. We will consider the effect on the node $k$ in the completed
quivers $\bar Q$ and $\bar Q'$. Given our ordering, suppose we draw
the nodes such that the labels increase from left to right. We then
have the following types of arrows associated to node $k$:
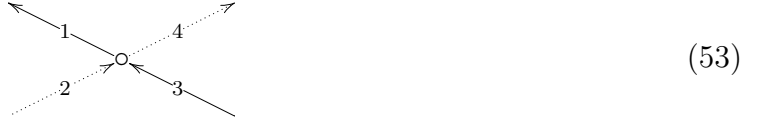
\begin{equation}
\begin{xy} <1.5mm,0mm>:
  (0,0)*{\circ}="a",
  (-10,5)="b",(-10,-5)="c",(10,-5)="d",(10,5)="e",
  \ar@{<-}|1 "b";"a"
  \ar@{.>}|2 "c";"a"
  \ar@{->}|3 "d";"a"
  \ar@{<.}|4 "e";"a"
\end{xy}  \label{eq:s1}
\end{equation}
Arrow 1 is associated to a $n_{kj}=1$ and so (\ref{eq:flip}) tells us
that in $\bar Q'$ this will be reversed. Arrow 2 was associated to an
$\Ext^2(L_k,L_j)$. According to (\ref{eq:tshf}), this will become
$\Ext^1(L_k',L_j')$. That is, this arrow will flip and become
solid. Arrow 3 will, by (\ref{eq:tshf}), become associated to an
$\Ext^2(L_j',L_k')$ so will flip and become dotted. Finally arrow 4
was associated to $\Ext^2(L_j,L_k)$ and so becomes an
$\Ext^3(L_j',L_k')$. As we saw in section \ref{ss:delP}, $\Ext^3$'s
lead to tachyons and destroy our interpretation of the quiver. Thus the
arrows in (\ref{eq:s1}) for $\bar Q$, become, in $\bar Q'$:
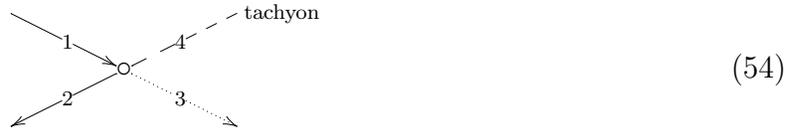
\begin{equation}
\begin{xy} <1.5mm,0mm>:
  (0,0)*{\circ}="a",
  (-10,5)="b",(-10,-5)="c",(10,-5)="d",(10,5)="e",
  (14,5)*{\hbox{\scriptsize tachyon}},
  \ar@{->}|1 "b";"a"
  \ar@{<-}|2 "c";"a"
  \ar@{<.}|3 "d";"a"
  \ar@{--}|4 "e";"a" 
\end{xy}  \label{eq:s2}
\end{equation}
So, if we impose the condition that there are no arrows of type 4, the
tilting transformation takes a valid gauge quiver to another gauge
quiver with no $\Ext^3$'s associated to node $k$ and reverses all the
arrows associated with node $k$. Furthermore, from (\ref{eq:Ni}), it
is easy to compute
\begin{equation}
\begin{split}
N_j' &= N_j,\quad j\neq k\\
N_k' &= \sum_i n_{ki}N_i - N_k.
\end{split}
\end{equation}
This transformation of the gauge groups together with the reversal of
all arrows associated to node $k$ is ``Seiberg duality at node $k$.''
Even though insisting that there are no arrows of type 4 is sufficient
to remove $\Ext^3$'s from node $k$ in the new quiver, there is no
guarantee that we don't induce new $\Ext^3$'s elsewhere in the
quiver. This tends not to happen in simple examples but we will see an
example of this occurrence in section \ref{ss:rem}.

Note that the ordering of the nodes in (\ref{eq:s2}) is now broken, as
arrow 1 points the wrong way. To regain the ordering, node $k$ must be
moved to the left of all nodes $j$ for which $n_{kj}\neq0$. This ties
in with the language of mutations as we discuss in section
\ref{ss:mut}.

Imposing the condition that there are no arrows of type 4
implies that there must be arrows of type 1 from (\ref{eq:anom1}) and
thus $n_{kj}\neq0$ for some $j$. This latter condition, which was
assumed at the start of this section, can therefore be subsumed by the
condition that there are no arrows of type 4.

The arrows in the quiver away from node $k$ will also be rearranged by
this tilting transformation and computing the $\Ext^1$'s and
$\Ext^2$'s for this transformation is straight-forward using the above
techniques. Note that we never need to compute the superpotential in
order to perform the Seiberg duality.

Unlike what one might expect from Seiberg duality, the functor
$\Psi_L$ is not its own inverse. Indeed, if we were to apply $\Psi_L$
again, then the arrows of type 3 would be tachyonic. Instead we define
another functor $\Psi_R$ given by the tilting complex
\begin{equation}
  T_R = \bigoplus_{j=0}^{n-1} P''_j,  \label{eq:TS2}
\end{equation}
where
\begin{equation}
\begin{split}
  P''_j &= P_j\quad\hbox{for $j\neq k$,}\\
  P''_k &= \bigl(\xymatrix@1{P_k
     \ar[r]&\poso{\bigoplus_j P_j^{\oplus n_{jk}}}}\bigr).
\end{split}
\end{equation}
It follows that $\Psi_R(L_k)=L_k''[-1]$ and, if $j\neq k$,
$\Psi_R(L_j)$ is determined by the short exact sequence
\begin{equation}
\xymatrix@1{
0\ar[r]&L_j''\ar[r]&\Psi_R(L_j)\ar[r]&L_k''^{\oplus n_{jk}}\ar[r]&0.
}
\end{equation}
Applying this transformation to the quiver piece (\ref{eq:s1}) then
yields
\begin{equation}
\begin{xy} <1.5mm,0mm>:
  (0,0)*{\circ}="a",
  (-10,5)="b",(-10,-5)="c",(10,-5)="d",(10,5)="e",
  (-14,-5)*{\hbox{\scriptsize tachyon}},
  \ar@{.>}|1 "b";"a"
  \ar@{--}|2 "c";"a"
  \ar@{<-}|3 "d";"a"
  \ar@{->}|4 "e";"a" 
\end{xy}  \label{eq:s3}
\end{equation}
Thus the arrows are reversed if there are no arrows of type
2. We also have
\begin{equation}
\begin{split}
N_j'' &= N_j,\quad j\neq k\\
N_k'' &= \sum_i n_{ik}N_i - N_k.
\end{split}
\end{equation}
Note that, if the anomaly condition (\ref{eq:anom1}) is satisfied, and
if there are no arrows of type 2 or 4, then $N_k''=N_k'$, and $\Psi_L$
and $\Psi_R$ both equally generate Seiberg duality.

If we apply $\Psi_L$ and then $\Psi_R$ then
\begin{equation}
\begin{split}
  P''_k &= \bigl(\xymatrix@1{P_k'
     \ar[r]&\poso{\bigoplus_j P_j'^{\oplus n'_{jk}}}}\bigr)\\
  &= \Cone\bigl(P_k' \to \bigoplus_j P_j'^{\oplus n'_{jk}}\bigr)\\
  &= \Cone\Bigl(\Cone\bigl(\bigoplus_j P_j^{\oplus n_{kj}}\to P_k\bigr))[-1] 
         \to \bigoplus_j P_j^{\oplus n_{kj}}\Bigr)\\
  &= P_k,
\end{split}
\end{equation}
and so $\Psi_R\circ\Psi_L$ is the identity. Similarly
$\Psi_L\circ\Psi_R$ is the identity too. Therefore, even though Seiberg is
na\"\i vely its own inverse, in the derived category picture it is
given by two functors which are each other's inverse.

In summary, the effect of the tilting transformation $\Psi_L$ for
Seiberg duality is to replace $L_k$ by $\Psi_L(L_k)=L_k'[1]$. Coarsely
speaking, we replace $L_k$ by its ``anti-brane.'' The remaining $L_j$'s
are replaced by bound states $L_j'\bind L_k'^{\oplus n_{kj}}$ or, more
precisely $\Cone(L_j'[-1]\to L_k'^{\oplus n_{kj}})$. This is
reminiscent of descriptions for Seiberg duality elsewhere (see
\cite{HW:3d,CFIKV:} for example) but we believe that the derived
category gives a much more precise picture as indicated in
\cite{BD:tilt}. 

The tilting transformation $\Psi_R$ also replaces
$L_k$ by an ``anti-brane,'' but this time the shift is the ``other
way,'' as $\Psi_R(L_k) = L_k''[-1]$. The bound states also use the open
strings in the opposite direction to $\Psi_L$. That is, $L_j$ is
replaced by $\Cone(L_k''^{\oplus n_{jk}}[-1]\to L_j'')$.  By insisting
on analyzing this problem using simplistic notions of anti-branes one
would fail to distinguish between these two transformations.

\subsection{Mutations} \label{ss:mut}

The functors $\Psi_L$ and $\Psi_R$ are equivalent
to specific left and right {\em mutations\/} of the exceptional
collection which makes contact with the work of \cite{CFIKV:,FHHI:quiv}. 

Suppose $\{\cF_0,\ldots,\cF_{n-1}\}$ corresponds to an exceptional
collection of sheaves. In derived category language the left-mutation
of $\cF_k$ through $\cF_{k-1}$ is defined by
\begin{equation}
  \mathsf{L}_{\cF_{k-1}}(\cF_k) =
     \Cone\bigl(\hom(\cF_{k-1},\cF_k)\otimes\cF_{k-1}\to\cF_k\bigr)[-1].
\end{equation}
Similarly, we define a right mutation:
\begin{equation}
  \mathsf{R}_{\cF_{k+1}}(\cF_k) =
     \Cone\bigl(\cF_k\to\hom(\cF_{k},\cF_{k+1})^*\otimes\cF_{k+1}\bigr).
\end{equation}
Now suppose $\cF_0,\ldots,\cF_{n-1}$ corresponds to a strongly
exceptional collection of sheaves. Choose a number $k$ and assume we
have numbered the collection so that there is another number $l<k$
such that $\Hom(\cF_j,\cF_k)$ is nonzero if and only if $l\leq j\leq
k$. Then the functor $\Psi_L$ is equivalent to mutating $\cF_k$
leftwards though members $k-1$ to $l$. $\Psi_R$ is similarly
constructed by right mutations.

In \cite{KO:excdP} it was shown that a left or right mutation in an
exceptional collection of sheaves on a del Pezzo surface yields a
complex with only one nonzero entry. Thus a mutation always yields 
an object of the form $\cE[n]$ for some sheaf $\cE$ and some integer
$n$. The value of $n$ is important (and distinguishes between
``division,'' ``extension,'' and ``recoil'' in mutation language
\cite{Gorod:Dhel}). 

As a deceptively simple example consider $\{\O,\O(1),\O(2)\}$ as an
exceptional collection on $\P^2$. We may reverse the order by mutating
$\O(2)$ to the left twice and $\O(1)$ to the left once to obtain the
exceptional collection $\{\O(-1),\Omega(1),\O\}$. In this case, no
shifts are involved. Now consider
$\{\O,\O(C_1),\O(H),\O(2H)\}$ as an exceptional collection on $\dP1$. 
Mutating the second entry to the left
yields $\{\O_{C_1}(-1)[-1],\O,\O(H),\O(2H)\}$. The first entry is now
shifted.

It is interesting to note (as effectively observed in \cite{Bon:quiv}
and used in \cite{HW:dib}) that the projective resolution of the
$L_k$'s can be phrased in terms of mutations. Up to shifts, one
mutates $L_k$ leftwards through all the members to the left. Carefully
following through the derived category computation one finds that
\begin{equation}
  L_k =
  \mathsf{L}_{P_0}\mathsf{L}_{P_1}\ldots\mathsf{L}_{P_{k-1}}P_k[k].
\end{equation}

It follows that all the fractional branes can be expressed as sheaves
on the del Pezzo surface shifted by some integer.  For example, the
fractional branes on $\P^2$ are $\O$, $\Omega(1)[1]$ and
$\O(-1)[2]$. Again we wish to emphasize the importance of taking the
shifts into account. If one were to merely assert (as is commonly
done) that the fractional branes were $\O$, anti-$\Omega(1)$, and
$\O(-1)$, we know of no systematic computation that would correctly
identify the spectrum of massless open strings between these
D-branes. As explained in \cite{me:TASI-D}, the shifts are also
important in seeing the quantum $\Z_3$ symmetry that arises in this
particular example.

It is not true that a mutation necessarily yields a tilting and thus a
Seiberg duality. The problem is that a strongly exceptional collection
may be mutated into a non-strongly exceptional collection. This
happened in section \ref{ss:refl} if $T_L$ or $T_R$ was not
admissible.  A similar failure can happen if $\cF_k$ is left-mutated
through only some, but not all, of the $\cF_j$'s with
$\Hom(\cF_j,\cF_k)\neq0$. For example, consider again the collection
$\{\O,\O(C),\O(H),\O(2H)\}$ on a $\dP1$. Mutating $\O(H)$ to the
left of $\O(C)$ results in an exceptional collection which is not
strong.  We therefore do not have a quiver interpretation of this
process. Mutating $\O(H)$ to the left of $\O(C)$ {\em and\/} $\O$
results in Seiberg duality on node $v_2$ in diagram (\ref{eq:dP1X}).

Of course many mutations, even if they are a tilting equivalence, will
produce $\Ext^3$'s and so will also not correspond to Seiberg
duality. Given an exceptional collection, the sequence of mutations
which reverses the order will typically produce $\Ext^3$'s. As we saw
above, such a sequence of mutations converts from the $P_i$ basis to the
$L_i$ basis. Therefore it is {\em not\/} generally true that these bases are
Seiberg dual to each other. 

The fact that an exceptional collection generates a ``helix''
\cite{Bon:quiv} implies that a sequence of tilting transformations
$\Psi_L$ will result in the identity transform. Given an exceptional
collection $\{\cF_0,\ldots,\cF_{n-1}\}$ on a del Pezzo surface we may
mutate $\cF_{n-1}$ leftwards through all $n-1$ elements on its left to
produce a new element
\begin{equation}
  \cF_{-1} = \cF_{n-1}\otimes K[3-n],
\end{equation}
where $K$ is the canonical sheaf. Similarly we may now mutate
$\cF_{n-2}$ all the way to the left. Continuing this process until we
have mutated $\cF_0$ to the left, we end up with a new exceptional
collection identical to the original one except that each element has
been tensored with $K[3-n]$. This tensoring has no effect on the
corresponding quiver and so we must return to the original quiver
after all these transformations. There is, of course, a similar
sequence of $\Psi_R$'s that also produces the identity transform.

\begin{figure}
\begin{equation}
\begin{xy} <0.5mm,0mm>:
  (0,0)*\xybox{
  (0,0)*{\circ}="a",(20,0)*{\circ}="b",(40,0)*{\circ}="c",(60,0)*{\circ}="d",
  \ar@{-}|*\dir{<}"a";"b"
  \ar@{-}@/_1mm/|*\dir{<}"b";"c"
  \ar@{-}|*\dir{<}"b";"c"
  \ar@{-}@/_3mm/|*\dir{<}"a";"c"
  \ar@{-}@/_1mm/|*\dir{<}"c";"d"
  \ar@{-}|*\dir{<}"c";"d"
  \ar@{-}@/^1mm/|*\dir{<}"c";"d"
  \ar@{.}@/^7mm/|*\dir{>}"a";"d"
  \ar@{.}@/^6mm/|*\dir{>}"a";"d"
  \ar@{.}@/^4mm/|*\dir{>}"b";"d"
  },
  (90,0)*\xybox{
  (0,0)*{\circ}="a",(20,0)*{\circ}="b",(40,0)*{\circ}="c",(60,0)*{\circ}="d",
  (40,-5)*{\scriptstyle 2}*\frm<1.5mm>{o},
  \ar@{-}|*\dir{<}"a";"b"
  \ar@{-}|*\dir{<}"b";"c"
  \ar@{-}@/_1mm/|*\dir{<}"c";"d"
  \ar@{-}|*\dir{<}"c";"d"
  \ar@{-}@/^1mm/|*\dir{<}"c";"d"
  \ar@{-}@/_2mm/|*\dir{<}"a";"c"
  \ar@{-}@/_3mm/|*\dir{<}"a";"c"
  \ar@{.}@/^6mm/|*\dir{>}^{\times 5}"a";"d"
  \ar@{.}@/^4mm/|*\dir{>}"b";"d"
  },
  (180,0)*\xybox{
  (0,0)*{\circ}="a",(20,0)*{\circ}="b",(40,0)*{\circ}="c",(60,0)*{\circ}="d",
  \ar@{-}|*\dir{<}"a";"b"
  \ar@{-}@/^1mm/|*\dir{<}"a";"b"
  \ar@{-}|*\dir{<}"b";"c"
  \ar@{-}|*\dir{<}"c";"d"
  \ar@{-}@/_1mm/|*\dir{<}"c";"d"
  \ar@{-}@/_3mm/|*\dir{<}"a";"c"
  \ar@{-}@/^3mm/|*\dir{<}"b";"d"
  \ar@{.}@/_6mm/|*\dir{>}_{\times 3}"a";"d"
  },
  (0,-40)*\xybox{
  (0,0)*{\circ}="a",(20,0)*{\circ}="b",(40,0)*{\circ}="c",(60,0)*{\circ}="d",
  (20,5)*{\scriptstyle 2}*\frm<1.5mm>{o},
  \ar@{-}@/^1mm/|*\dir{<}"a";"b"
  \ar@{-}|*\dir{<}"a";"b"
  \ar@{-}@/_1mm/|*\dir{<}"a";"b"
  \ar@{-}|*\dir{<}"b";"c"
  \ar@{-}|*\dir{<}"c";"d"
  \ar@{-}@/^3mm/|*\dir{<}"b";"d"
  \ar@{-}@/^2mm/|*\dir{<}"b";"d"
  \ar@{.}@/_4mm/|*\dir{>}"a";"c"
  \ar@{.}@/_6mm/|*\dir{>}_{\times 5}"a";"d"
  },
  (90,-40)*\xybox{
  (0,0)*{\circ}="a",(20,0)*{\circ}="b",(40,0)*{\circ}="c",(60,0)*{\circ}="d",
  \ar@{-}@/^1mm/|*\dir{<}"a";"b"
  \ar@{-}|*\dir{<}"a";"b"
  \ar@{-}@/_1mm/|*\dir{<}"a";"b"
  \ar@{-}@/_1mm/|*\dir{<}"b";"c"
  \ar@{-}@/^1mm/|*\dir{<}"b";"c"
  \ar@{-}|*\dir{<}"c";"d"
  \ar@{-}@/^2mm/|*\dir{<}"b";"d"
  \ar@{.}@/_4mm/|*\dir{>}"a";"c"
  \ar@{.}@/_7mm/|*\dir{>}"a";"d"
  \ar@{.}@/_6mm/|*\dir{>}"a";"d"
  },
  (180,-40)*\xybox{
  (0,0)*{\circ}="a",(20,0)*{\circ}="b",(40,0)*{\circ}="c",(60,0)*{\circ}="d",
  \ar@{-}|*\dir{<}"a";"b"
  \ar@{-}@/^1mm/|*\dir{<}"a";"b"
  \ar@{-}|*\dir{<}"b";"c"
  \ar@{-}|*\dir{<}"c";"d"
  \ar@{-}@/_1mm/|*\dir{<}"c";"d"
  \ar@{-}@/_3mm/|*\dir{<}"a";"c"
  \ar@{-}@/^3mm/|*\dir{<}"b";"d"
  \ar@{.}@/_6mm/|*\dir{>}_{\times 3}"a";"d"
  },
  (0,-80)*\xybox{
  (0,0)*{\circ}="a",(20,0)*{\circ}="b",(40,0)*{\circ}="c",(60,0)*{\circ}="d",
  \ar@{-}@/^1mm/|*\dir{<}"a";"b"
  \ar@{-}@/_1mm/|*\dir{<}"a";"b"
  \ar@{-}@/^1mm/|*\dir{<}"b";"c"
  \ar@{-}|*\dir{<}"b";"c"
  \ar@{-}@/_1mm/|*\dir{<}"b";"c"
  \ar@{-}@/_1mm/|*\dir{<}"c";"d"
  \ar@{-}@/^1mm/|*\dir{<}"c";"d"
  \ar@{.}@/^6mm/|*\dir{>}"a";"d"
  \ar@{.}@/^4mm/|*\dir{>}"b";"d"
  \ar@{.}@/_4mm/|*\dir{>}"a";"c"
  },
  (90,-80)*\xybox{
  (0,0)*{\circ}="a",(20,0)*{\circ}="b",(40,0)*{\circ}="c",(60,0)*{\circ}="d",
  \ar@{-}|*\dir{<}"a";"b"
  \ar@{-}@/^1mm/|*\dir{<}"a";"b"
  \ar@{-}|*\dir{<}"b";"c"
  \ar@{-}|*\dir{<}"c";"d"
  \ar@{-}@/_1mm/|*\dir{<}"c";"d"
  \ar@{-}@/_3mm/|*\dir{<}"a";"c"
  \ar@{-}@/^3mm/|*\dir{<}"b";"d"
  \ar@{.}@/_6mm/|*\dir{>}_{\times 3}"a";"d"
  },
  (180,-80)*\xybox{
  (0,0)*{\circ}="a",(20,0)*{\circ}="b",(40,0)*{\circ}="c",(60,0)*{\circ}="d",
  \ar@{-}|*\dir{<}"a";"b"
  \ar@{-}@/_1mm/|*\dir{<}"b";"c"
  \ar@{-}|*\dir{<}"b";"c"
  \ar@{-}@/_3mm/|*\dir{<}"a";"c"
  \ar@{-}@/_1mm/|*\dir{<}"c";"d"
  \ar@{-}|*\dir{<}"c";"d"
  \ar@{-}@/^1mm/|*\dir{<}"c";"d"
  \ar@{.}@/^7mm/|*\dir{>}"a";"d"
  \ar@{.}@/^6mm/|*\dir{>}"a";"d"
  \ar@{.}@/^4mm/|*\dir{>}"b";"d"
  },
 \ar@{->}^{3}(70,0);(80,0)
 \ar@{->}^{2}(160,0);(170,0)
 \ar@{->}^{3}(250,0);(260,0)
 \ar@{->}^{1}(70,-40);(80,-40)
 \ar@{->}^{3}(160,-40);(170,-40)
 \ar@{->}^{1}(250,-40);(260,-40)
 \ar@{->}^{3}(70,-80);(80,-80)
 \ar@{->}^{2}(160,-80);(170,-80)
\end{xy} 
\end{equation}
\caption{A sequence of $\Psi_L$ transforms equivalent to the identity
  for a $\dP1$.}  \label{fig:helix}
\end{figure}
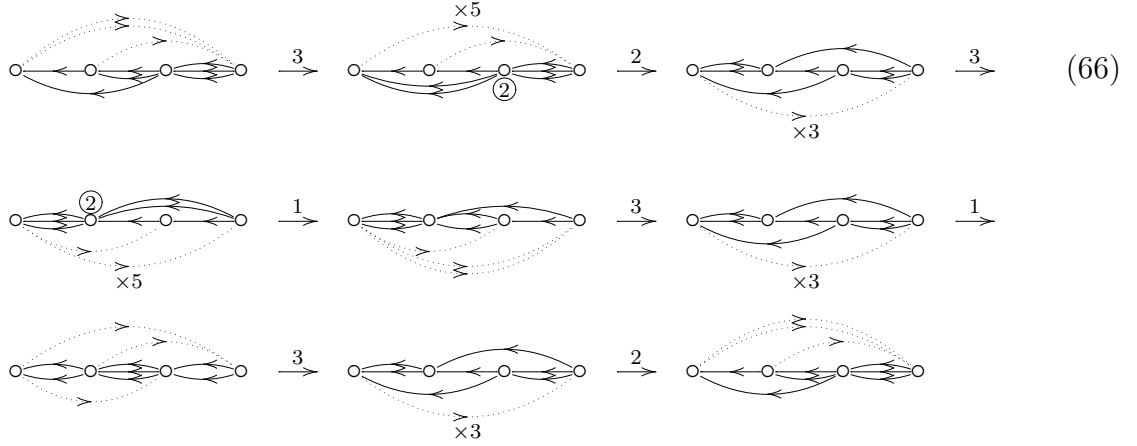

As an example, consider the quiver given by (\ref{eq:dP1X}) for a
$\dP1$. The sequence of tilting transforms given by $\Psi_L$
associated to the above helix is shown in figure \ref{fig:helix}.  The
node (counting from left to right starting at 0) to which Seiberg
duality is applied is shown over the arrows between the diagrams. A
number in a circle gives that value of $N_i$ (i.e., the rank of the
gauge group) for that node (the default being 1). After applying the
tilting transformation we have reordered the nodes to get back to an
ordered graph. This is exactly the reordering process familiar in
mutations.  All our diagrams will use this reordering from now on.
Note that in figure \ref{fig:helix} we have certain duplications
appearing in the sequence, but this is not a general feature in more
complicated examples.

\subsection{Removing $\Ext^3$'s}  \label{ss:rem}

As well as producing a Seiberg duality, we may also use the tilting of
section \ref{ss:refl} to remove unwanted $\Ext^3$'s from quivers. For
example, in section \ref{ss:delP} we considered the quiver
(\ref{eq:dP4}) associated to a $\dP4$ for which $\Ext^3(L_6,L_0)$ is
nonzero. By applying $\Psi_R$ to node 0 we convert this to an $\Ext^2$
yielding the quiver
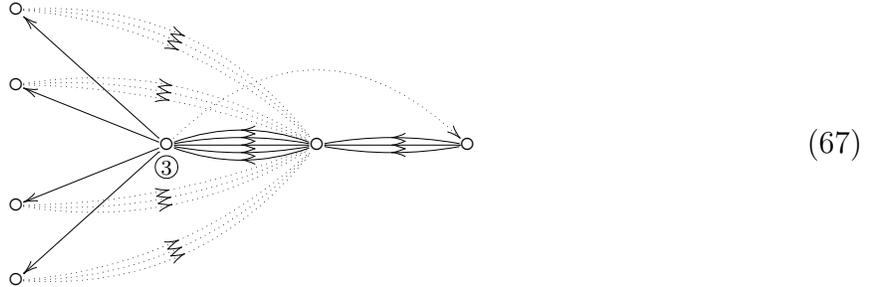
\begin{equation}
\begin{xy} <1.0mm,0mm>:
  (20,0)*{\circ}="a",
  (0,18)*{\circ}="b1",
  (0,8)*{\circ}="b2",
  (0,-8)*{\circ}="b3",
  (0,-18)*{\circ}="b4",
  (40,0)*{\circ}="c",(60,0)*{\circ}="d",
  (20,-3)*{\scriptstyle 3}*\frm<1.5mm>{o},
  \ar@{->}"a";"b1"
  \ar@{->}"a";"b2"
  \ar@{->}"a";"b3"
  \ar@{->}"a";"b4"
  \ar@{.}@/^6mm/|*\dir{>}"b1";"c"
  \ar@{.}@/^5mm/|*\dir{>}"b1";"c"
  \ar@{.}@/^4mm/|*\dir{>}"b1";"c"
  \ar@{.}@/^4mm/|*\dir{>}"b2";"c"
  \ar@{.}@/^3mm/|*\dir{>}"b2";"c"
  \ar@{.}@/^2mm/|*\dir{>}"b2";"c"
  \ar@{.}@/_2mm/|*\dir{>}"b3";"c"
  \ar@{.}@/_3mm/|*\dir{>}"b3";"c"
  \ar@{.}@/_4mm/|*\dir{>}"b3";"c"
  \ar@{.}@/_4mm/|*\dir{>}"b4";"c"
  \ar@{.}@/_5mm/|*\dir{>}"b4";"c"
  \ar@{.}@/_6mm/|*\dir{>}"b4";"c"
  \ar@{-}@/^1mm/|*\dir{<}"c";"d"
  \ar@{-}|*\dir{<}"c";"d"
  \ar@{-}@/_1mm/|*\dir{<}"c";"d"
  \ar@{.>}@/^10mm/"a";"d"
  \ar@{-}@/_0mm/|*\dir{<}"a";"c"
  \ar@{-}@/_1mm/|*\dir{<}"a";"c"
  \ar@{-}@/_2mm/|*\dir{<}"a";"c"
  \ar@{-}@/^1mm/|*\dir{<}"a";"c"
  \ar@{-}@/^2mm/|*\dir{<}"a";"c"
\end{xy}  \label{eq:dP4a}
\end{equation}

Alternatively we could apply $\Psi_L$ to node 6 to produce another
valid quiver:
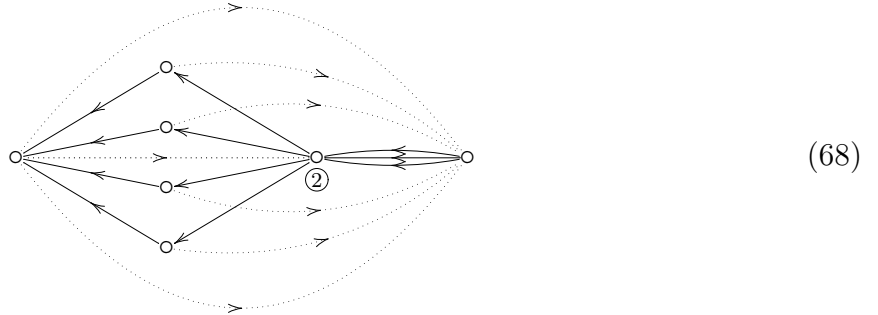
\begin{equation}
\begin{xy} <1.0mm,0mm>:
  (0,0)*{\circ}="a",
  (20,12)*{\circ}="b1",
  (20,4)*{\circ}="b2",
  (20,-4)*{\circ}="b3",
  (20,-12)*{\circ}="b4",
  (60,0)*{\circ}="c",(40,0)*{\circ}="d",
  (40,-3)*{\scriptstyle 2}*\frm<1.5mm>{o},
  \ar@{-}|*\dir{<}"a";"b1"
  \ar@{-}|*\dir{<}"a";"b2"
  \ar@{-}|*\dir{<}"a";"b3"
  \ar@{-}|*\dir{<}"a";"b4"
  \ar@{.}@/^5mm/|*\dir{>}"b1";"c"
  \ar@{.}@/^5mm/|*\dir{>}"b2";"c"
  \ar@{.}@/_5mm/|*\dir{>}"b3";"c"
  \ar@{.}@/_5mm/|*\dir{>}"b4";"c"
  \ar@{-}@/^1mm/|*\dir{>}"c";"d"
  \ar@{-}|*\dir{>}"c";"d"
  \ar@{-}@/_1mm/|*\dir{>}"c";"d"
  \ar@{.}|*\dir{>}"a";"d"
  \ar@{.}@/_20mm/|*\dir{>}"a";"c"
  \ar@{.}@/^20mm/|*\dir{>}"a";"c"
  \ar@{->}"d";"b1"
  \ar@{->}"d";"b2"
  \ar@{->}"d";"b3"
  \ar@{->}"d";"b4"
\end{xy}  \label{eq:dP4b}
\end{equation}

The nonzero $\Ext^3$ in the original quiver induced a tachyon, and so
one would expect the field theory to seek out a true vacuum by giving
an expectation value to this mode. The two quivers above are not
expected to be the new vacuum. This is because, as explained in
section \ref{ss:refl}, the tiltings are obtained by giving expectation
values to other arrows in the quiver (corresponding to massless modes
for our chosen point in moduli space). It would be interesting to
determine the quiver that results from the tachyon field acquiring a
vev.

It is possible that applying $\Psi_L$ or $\Psi_R$ can eliminate
unwanted $\Ext^3$'s from node $k$, but this procedure can induce new
$\Ext^3$'s elsewhere. As an example, consider the exceptional
collection $\{\O,\O(C_1),\ldots,\O(C_8),\O(H),\O(2H)\}$ on
a $\dP8$. This has a quiver
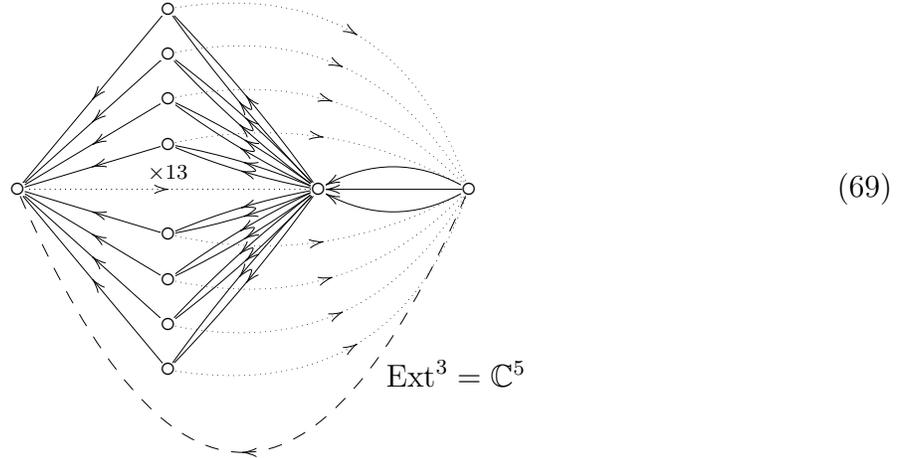
\begin{equation}
\begin{xy} <1.0mm,0mm>:
  (0,0)*{\circ}="a",
  (20,24)*{\circ}="b1",
  (20,18)*{\circ}="b2",
  (20,12)*{\circ}="b3",
  (20,6)*{\circ}="b4",
  (20,-6)*{\circ}="b5",
  (20,-12)*{\circ}="b6",
  (20,-18)*{\circ}="b7",
  (20,-24)*{\circ}="b8",
  (40,0)*{\circ}="c",(60,0)*{\circ}="d",
  \ar@{-}|*\dir{<}"a";"b1"
  \ar@{-}|*\dir{<}"a";"b2"
  \ar@{-}|*\dir{<}"a";"b3"
  \ar@{-}|*\dir{<}"a";"b4"
  \ar@{-}|*\dir{<}"a";"b5"
  \ar@{-}|*\dir{<}"a";"b6"
  \ar@{-}|*\dir{<}"a";"b7"
  \ar@{-}|*\dir{<}"a";"b8"
  \ar@{-}@/^0.6mm/|*\dir{<}"b1";"c"
  \ar@{-}@/_0.6mm/|*\dir{<}"b1";"c"
  \ar@{-}@/^0.6mm/|*\dir{<}"b2";"c"
  \ar@{-}@/_0.6mm/|*\dir{<}"b2";"c"
  \ar@{-}@/^0.6mm/|*\dir{<}"b3";"c"
  \ar@{-}@/_0.6mm/|*\dir{<}"b3";"c"
  \ar@{-}@/^0.6mm/|*\dir{<}"b4";"c"
  \ar@{-}@/_0.6mm/|*\dir{<}"b4";"c"
  \ar@{-}@/^0.6mm/|*\dir{<}"b5";"c"
  \ar@{-}@/_0.6mm/|*\dir{<}"b5";"c"
  \ar@{-}@/^0.6mm/|*\dir{<}"b6";"c"
  \ar@{-}@/_0.6mm/|*\dir{<}"b6";"c"
  \ar@{-}@/^0.6mm/|*\dir{<}"b7";"c"
  \ar@{-}@/_0.6mm/|*\dir{<}"b7";"c"
  \ar@{-}@/^0.6mm/|*\dir{<}"b8";"c"
  \ar@{-}@/_0.6mm/|*\dir{<}"b8";"c"
  \ar@{<-}@/^3mm/"c";"d"
  \ar@{<-}"c";"d"
  \ar@{<-}@/_3mm/"c";"d"
  \ar@{.}|*\dir{>}^*{\scriptstyle \times 13}"a";"c"
  \ar@{.}@/^10mm/|*\dir{>}"b1";"d"
  \ar@{.}@/^8mm/|*\dir{>}"b2";"d"
  \ar@{.}@/^6mm/|*\dir{>}"b3";"d"
  \ar@{.}@/^4mm/|*\dir{>}"b4";"d"
  \ar@{.}@/_4mm/|*\dir{>}"b5";"d"
  \ar@{.}@/_6mm/|*\dir{>}"b6";"d"
  \ar@{.}@/_8mm/|*\dir{>}"b7";"d"
  \ar@{.}@/_10mm/|*\dir{>}"b8";"d"
  \ar@{--}@/_35mm/|*\dir{<}_(0.8)*{\Ext^3=\C^5}"a";"d"
\end{xy}  \label{eq:dP8}
\end{equation}
Applying $\Psi_L$ to the right-most node we may remove the problematic
$\Ext^3$. The resulting quiver is now
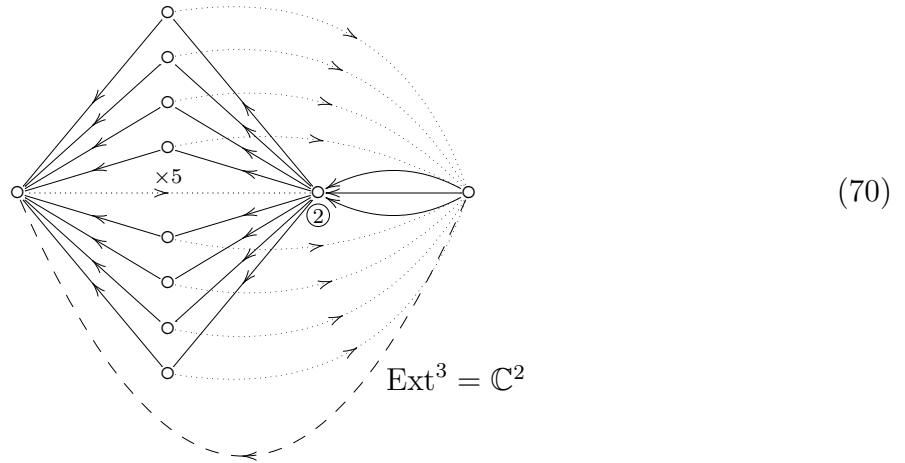
\begin{equation}
\begin{xy} <1.0mm,0mm>:
  (0,0)*{\circ}="a",
  (20,24)*{\circ}="b1",
  (20,18)*{\circ}="b2",
  (20,12)*{\circ}="b3",
  (20,6)*{\circ}="b4",
  (20,-6)*{\circ}="b5",
  (20,-12)*{\circ}="b6",
  (20,-18)*{\circ}="b7",
  (20,-24)*{\circ}="b8",
  (40,0)*{\circ}="c",(60,0)*{\circ}="d",
  (40,-3)*{\scriptstyle 2}*\frm<1.5mm>{o},
  \ar@{-}|*\dir{<}"a";"b1"
  \ar@{-}|*\dir{<}"a";"b2"
  \ar@{-}|*\dir{<}"a";"b3"
  \ar@{-}|*\dir{<}"a";"b4"
  \ar@{-}|*\dir{<}"a";"b5"
  \ar@{-}|*\dir{<}"a";"b6"
  \ar@{-}|*\dir{<}"a";"b7"
  \ar@{-}|*\dir{<}"a";"b8"
  \ar@{-}|*\dir{<}"b1";"c"
  \ar@{-}|*\dir{<}"b2";"c"
  \ar@{-}|*\dir{<}"b3";"c"
  \ar@{-}|*\dir{<}"b4";"c"
  \ar@{-}|*\dir{<}"b5";"c"
  \ar@{-}|*\dir{<}"b6";"c"
  \ar@{-}|*\dir{<}"b7";"c"
  \ar@{-}|*\dir{<}"b8";"c"
  \ar@{<-}@/^3mm/"c";"d"
  \ar@{<-}"c";"d"
  \ar@{<-}@/_3mm/"c";"d"
  \ar@{.}|*\dir{>}^*{\scriptstyle \times 5}"a";"c"
  \ar@{.}@/^10mm/|*\dir{>}"b1";"d"
  \ar@{.}@/^8mm/|*\dir{>}"b2";"d"
  \ar@{.}@/^6mm/|*\dir{>}"b3";"d"
  \ar@{.}@/^4mm/|*\dir{>}"b4";"d"
  \ar@{.}@/_4mm/|*\dir{>}"b5";"d"
  \ar@{.}@/_6mm/|*\dir{>}"b6";"d"
  \ar@{.}@/_8mm/|*\dir{>}"b7";"d"
  \ar@{.}@/_10mm/|*\dir{>}"b8";"d"
  \ar@{--}@/_35mm/|*\dir{<}_(0.8)*{\Ext^3=\C^2}"a";"d"
\end{xy}  \label{eq:dP8a}
\end{equation}
Thus, we gained some new $\Ext^3$'s resulting in another invalid quiver.

\subsection{Some conjectures}  \label{ss:conj}

Having explored a large number of tilting transformations we would
like to make the following conjectures which seem to always hold. All of
these conjectures are essentially statements about linear algebra so
there may well be a simple proof that we have missed.

Let us say that $T_L$ is a ``valid'' tilting complex for a given node
if there are no arrows of type 4, and $T_R$ is valid if there are no
arrows of type 2.

\begin{conjecture}~
\begin{enumerate}
  \item All valid $T$'s are admissible (in the sense of section
\ref{ss:refl}).
  \item If $T_L$ and $T_R$ are both valid for a given node then they
  yield the same completed quiver (making no distinction between
  $\Ext^1$'s and $\Ext^2$'s).
  \item For any node, either $T_L$ or $T_R$ is valid (i.e., we never
  have arrows of type 2 and 4 on a particular node).
  \item A valid $T$ applied to a quiver free of $\Ext^3$'s will not
  induce any new $\Ext^3$'s.
\end{enumerate}
\end{conjecture}

Assuming these conjectures are correct, it makes Seiberg duality
straight-forward for quivers. We are free to apply duality to any node
and we obtain a unique valid result. 


\section{Conclusions}  \label{s:conc}

We have shown in detail that a 0-brane on a collapsed del Pezzo
surface decays marginally into a set of fractional branes in a way
described by a quiver. The B-type D-branes on this del Pezzo surface
are then described by the derived category of representations of this
quiver. This allowed us to describe a very large class of Seiberg
dualities using the language of tilting equivalences.

It is very important to distinguish between $\Ext^1$'s and $\Ext^3$'s
between D-branes. The former give massless bifundamental chiral
fields, while the latter give tachyons. Any analysis which only uses
only intersection pairings at the K-theory level (such as $\chi(A,B)$
or, in the mirror, intersection numbers of 3-cycles) will miss this
distinction. Fortunately the derived category provides a rigorous and
complete framework for understanding this problem.

The general program of trying to use tilting equivalences to analyze
Seiberg duality is frustrated by the appearance of
infinite-dimensional representations caused by oriented loops in the
quiver. In the case of del Pezzo surfaces this problem is avoided and
the tilting transformations become equivalent to mutations of
exceptional collections of sheaves. It may well be possible to deal
directly with the quivers containing loops, as was done in
\cite{Brn:tilt}. It may also provide some insight into the conjectures
of section \ref{ss:conj}.

The obvious extension of our work would be to consider {\em
generalized\/} del Pezzo surfaces and to consider more than one
such surface collapsing to the same point. This would provide a
general analysis of all singularities that can occur at a point which
can be blown-up by an exceptional divisor. This would, of course,
include all orbifolds but would provide a much larger class. It would
be interesting to see if representations of quiver path algebras would
continue to play a useful r\^ole in this wider problem. Some results
on weighted projective space along this line are discussed in
\cite{GL:wcp-q,Baer:quiv}.

We have had very little to say about the superpotential in this
paper. Indeed, it is quite interesting to note that the tilting
transformations allow one to compute Seiberg dualities without knowing
the superpotential. Having said that, the derived category of quiver
representations does provide all information about the superpotential
in the form of $A_\infty$ algebras along the lines of
\cite{HLW:Ainf}. Presumably this ties in with the work already done on
superpotentials in this subject \cite{CFIKV:,Wijn:dP}. The arena of
del Pezzo surfaces should provide many nice examples for studying the
$A_\infty$ structure.


\section*{Acknowledgments}

We wish to thank C.~Haase, A.~Hanany, D.~Morrison, and R.~Plesser for useful
conversations. The authors are supported in part by NSF grants
DMS-0074072 and DMS-0301476.


\end{document}